\documentclass{ptephy_v1}%%%%where ptephy_v1 is the template name
\usepackage{subfig}
\usepackage{amsmath}
\usepackage{graphics}
\usepackage{url}

\usepackage{color}

\def\edth{{\rlap{$\partial$}\raise0.3em\hbox{$-$}}}
\newcommand{\bea}{\begin{eqnarray}}
\newcommand{\eea}{\end{eqnarray}}

\newcommand{\msun}{{M}_{\odot}}
\newcommand{\rsun}{{R}_{\odot}}

\begin{document}

\title{Gravitational wave quasinormal mode from Population III massive black hole binaries
in various models of population synthesis}

%%%% To generate auto affiliation numbers please use \author{}\affil{} command

\author{\name{Tomoya Kinugawa}{1},
\name{Hiroyuki Nakano}{2}
and \name{Takashi Nakamura}{2}}

\address{${}^1$\affil{1}{Institute for Cosmic Ray Research, The University of Tokyo,
Chiba 277-8582, Japan}
\\
${}^2$\affil{2}{Department of Physics, Kyoto University, Kyoto 606-8502, Japan}
\\
}

%%%%%%%%%%%%%%%%%%%%%%%%%%%%%%%%%%%%%%%%
\begin{abstract}
Focusing on the remnant black holes after merging binary black holes,
we show that ringdown gravitational waves of Population III binary black holes mergers can be detected with
the rate of
$5.9-500~{\rm events~yr^{-1}}~({\rm SFR_p}/ (10^{-2.5}~M_\odot~{\rm yr^{-1}~Mpc^{-3}})) 
\cdot ({\rm [f_b/(1+f_b)]/0.33})$
for various parameters and functions.
This rate is estimated for the events with SNR$>8$
for the second generation gravitational wave detectors such as KAGRA.
Here, ${\rm SFR_p}$ and ${\rm f_b}$ are the peak value of the Population III star
formation rate and the fraction of binaries, respectively.
When we consider only the events with SNR$>35$,
the event rate becomes 
$0.046-4.21~{\rm events~yr^{-1}}~({\rm SFR_p}/ (10^{-2.5}~M_\odot~{\rm yr^{-1}~Mpc^{-3}})) 
\cdot ({\rm [f_b/(1+f_b)]/0.33})$.
This suggest that for remnant black hole's spin $q_f>0.95$
we have the event rate with SNR$>35$ less than
$0.037~{\rm events~yr^{-1}}~({\rm SFR_p}/ (10^{-2.5}~M_\odot~{\rm yr^{-1}~Mpc^{-3}})) 
\cdot ({\rm [f_b/(1+f_b)]/0.33})$,
while it is
$3-30~{\rm events~yr^{-1}}~({\rm SFR_p}/ (10^{-2.5}~M_\odot~{\rm yr^{-1}~Mpc^{-3}})) 
\cdot ({\rm [f_b/(1+f_b)]/0.33})$  
for the third generation detectors such as Einstein Telescope.
If we detect many Population III binary black holes merger,
it may be possible to constrain the Population III binary evolution paths
not only by the mass distribution but also by the spin distribution.

\end{abstract}
%%%%%%%%%%%%%%%%%%%%%%%%%%%%%%%%%%%%%%%%

\subjectindex{E31, E02, E01, E38}
%E01 Relativity
%E02 Gravitational waves
%E31 Black holes
%E38 Physics of strong fields

\maketitle

%%%%%%%%%%%%%%%%%%%%%%%%%%%%%%%%%%%%%%%%
\section{Introduction}
%%%%%%%%%%%%%%%%%%%%%%%%%%%%%%%%%%%%%%%%

The final part of gravitational waves (GWs) from merging binary black holes (BBHs)
is called the ringdown phase. When the remnant compact object is a black hole (BH),
this phase is described by quasinormal modes (QNMs) of the BH
(see. e.g., Ref.~\cite{Cardoso:2016rao}).
In general, the BH is expected as the Kerr spacetime~\cite{Kerr:1963ud},
\bea
 ds^2 &=& 
 - \left( 1 - \frac{2 M r}{\Sigma} \right)  dt^2 
 - \frac{4 M a r ~{\rm{sin}^2 \theta} }{\Sigma} dt d\phi 
 + \frac{\Sigma}{\Delta} dr^2
 \cr &&
 + \Sigma d\theta^2
 + \left( r^2 + a^2 + \frac{2 M a^2 r}{\Sigma} \sin^2 \theta \right)
 \sin^2 \theta d\phi^2 \,,
\eea
where $\Delta = r^2 - 2 M r + a^2$ and $\Sigma = r^2 + a^2 \cos^2 \theta$.
with mass $M$ and spin $a$.
The detection of QNM GWs is not only to give a precise estimation
of the BH's mass and spin, but also to test Einstein's general relativity
(see an extensive review~\cite{Berti:2015itd}).

In our previous paper~\cite{Kinugawa:2016mfs}, 
using the recent population synthesis results
of Population III (Pop III) massive BBHs~\cite{Kinugawa:2014zha,Kinugawa:2015nla},
we discussed the event rate of the QNM GWs
by  the second generation gravitational wave detectors
such as Advanced LIGO (aLIGO)~\cite{TheLIGOScientific:2014jea}, 
Advanced Virgo (AdV)~\cite{TheVirgo:2014hva}, 
and KAGRA~\cite{Somiya:2011np,Aso:2013eba}.
Since there are various parameters and functions in the population synthesis calculation,
we extensively examine BBH binary formations in this paper.
Then,  we calculate the remnant BH's mass $M_f$
and the non-dimensional spin $q_f=a_f/M_f$ via fitting formulas~\cite{Healy:2014yta},
and the event rate for each model.

This paper is organized as follows.
In Sec.~\ref{sec:PopIIIBPS}, we summarize our Pop III binary population synthesis calculation,
and prepare 10 models.
In Sec.~\ref{sec:MR_BBHR},
we show the mass ratio distributions of BBH remnants
for each model. The dependences on initial distribution functions,
binary parameters etc. are discussed there.
In Sec.~\ref{S_BBHR}, we obtain
the spin distributions of BBH remnants
for various parameters.
Using the results presented in the above sections,
we calculate the mass and spin distributions of the remnant BHs after merger
in Sec.~\ref{sec:R_ER}, and show the event rates  of QNMs for each mode in KAGRA.
Final section~\ref{sec:dis} is devoted to discussions.

%%%%%%%%%%%%%%%%%%%%%%%%%%%%%%%%%%%%%%%%
\section{Population III binary population synthesis calculation}\label{sec:PopIIIBPS}
%%%%%%%%%%%%%%%%%%%%%%%%%%%%%%%%%%%%%%%%

To estimate the detection rate of GWs from Pop III BBH mergers,
it is necessary to know how many Pop III binaries
become BBHs which merge within the Hubble time.
Here, we use the binary population synthesis method
which is the Monte Carlo simulation of binary evolutions.
The Pop III binary population synthesis
code~\cite{Kinugawa:2014zha, Kinugawa:2015nla, Kinugawa:2016mfs}
has been upgraded from the binary population synthesis code~\cite{Hurley:2002rf,hurleysite}
for Pop III binaries.
In this paper, we calculate the same models as Ref.~\cite{Kinugawa:2015nla}
using the same methods as Ref.~\cite{Kinugawa:2015nla}
in order to obtain the mass ratio distribution and the spin distribution.
In this section, we review the calculation method and models.
Note that in this paper, we do not consider the kick models and
the worst model discussed in Ref.~\cite{Kinugawa:2015nla} for simplicity,
because in these models, BBHs have misaligned spins and the final spins after merger
are too complex. 

First, we need to give initial conditions when a binary is born.
The initial conditions such as primary mass $M_1$,
mass ratio $M_2/M_1$ where $M_2$ is the secondary mass, separation $a$
and orbital eccentricity $e$ are decided by Monte Carlo method
with initial distribution functions such as the initial mass function (IMF),
the initial mass ratio function (IMRF),
the initial separation function (ISF),
and the initial eccentricity function (IEF).
For example, in our standard model, we use the flat IMF,
the flat IMRF, the logflat ISF and the IEF with a function of $\propto e$.
There is no observation of Pop III binary because they born at the early universe.
Thus, we do not know the initial distribution functions of Pop III binaries from the observation.
For the IMF, however, the recent simulations~\cite{Hirano:2013lba, Susa:2014moa}
may suggest the flat IMF. Therefore, we adapt the flat IMF.
For the other initial distribution functions, we adapt those of Pop I case
where Pop I star is solar like star.
The set of the above initial distribution functions is called as our standard model in this paper.

Second, we calculate the evolution of each star, 
and if the star satisfies a condition of binary interactions,
we evaluate the effects of binary interactions and change $M_1$, $M_2$, $a$ and $e$.
As the binary interactions, we treat the Roche lobe overflow (RLOF),
the common envelope (CE) phase, the tidal effect, the supernova effect,
and the gravitational radiation.
The RLOF is the stable mass transfer, 
while the unstable mass transfer becomes the CE phase
when the donor star is a giant.
Here, we need some parameters for the calculation of the RLOF and the CE phase.

In the case of the RLOF, we use the lose fraction $\beta$
of transfered stellar matter defined as
\begin{equation}\label{betaMT}
 \dot{M}_2 = - (1-\beta) \dot{M}_1 \,.
\end{equation}
where $\dot{M_2}$ is the mass accretion rate of a receiver star,
$\dot{M_1}$ is the mass loss rate of a donor star.
In our standard model, $\beta$ is determined by the Hurley's function~\cite{Hurley:2002rf}
which has been discussed for the Pop I case.
When the receiver star is in the main sequence phase or in the He-burning phase,
we assume that the accretion rate is described by
\begin{equation}\label{MT}
 \dot{M}_{2} = -{\rm{min}}\left(10\frac{\tau_{\dot{M}}}{\tau_{{\rm{KH,2}}}},1\right)\dot{M}_{1} \,,
\end{equation}
where $\tau_{\dot{M}}$ is the accretion time scale defined by
\begin{equation}
 \tau_{\dot{M}}\equiv\frac{M_2}{\dot{M}_1} \,,
\end{equation}
and the Kelvin-Helmholtz timescale $\tau_{\rm KH,2}$ is defined by
\begin{align}
 \tau_{\rm{KH,2}} = \frac{GM_2(M_2-M_{\rm{c},2})}{L_2R_2} \,.
\end{align}
Here, $M_2$, $M_{\rm{c},2}$, $L_2$ and $R_2$ are the mass,
the core mass, the luminosity and the radius of the receiver star, respectively.
When the receiver star is in the He-shell burning phase,
we assume that the receiver star can get all transfered matter from the donor star, i.e., 
\begin{equation}\label{MT2}
 \dot{M}_{2} = -\dot{M}_{1} \,.
\end{equation}
Although we use the $\beta$ function defined by Hurley et al.~\cite{Hurley:2002rf}
in our standard model, we also treat the accretion rate of the receiver star
described by the constant $\beta$ parameter.
This is because the accretion rate of the receiver star which is not a compact object, 
is not understood well. 
Furthermore, in our previous study~\cite{Kinugawa:2015nla},
we have shown that the Hurley fitting formula is consistent with $\beta=0$ in the Pop III binary case.
Thus, we also discuss the cases of $\beta=0.5$ and $\beta=1$.
It is noted that the stability of the mass transfer changes
if the mass transfer is nonconservative ($\beta>0$).
We use the criterion given in Ref.~\cite{Eggleton2011} as
\begin{align}
 \zeta_{\rm L}
 &=\frac{d{\rm{log}}R_{\rm L,1}}{d{\rm{log}}M_1}
 \cr
 &= \biggl[
 \left( 0.33+0.13 \frac{M_1}{M_2} \right) \left(1+\frac{M_1}{M_2}-\beta \frac{M_1}{M_2}\right)
 +(1-\beta) \left(\left(\frac{M_1}{M_2}\right)^2-1\right)
 \cr
 & \quad -\beta \frac{M_1}{M_2} \left. \biggr] \right/ \left(1+\frac{M_1}{M_2}\right) \,,
\end{align}   
where $M_1$ and $R_{\rm L,1}$ are the mass and the Roche lobe radius of the donor star.
If $\zeta_{\rm ad} = d\log R_{{\rm ad},1}/d\log M_1<\zeta_{\rm L}$
where $R_{\rm ad,1}$ is the radius of the donor star,
in the hydrostatic equilibrium of the donor star,
the binary starts a dynamically unstable mass transfer such as the CE phase. 
When the receiver star is a compact object such as a neutron star and a BH,
we always use $\beta=0$ and the upper limit of the accretion rate
is limited by the Eddington accretion rate defined by
\begin{align}
 \dot{M}_{\rm{Edd}}&=\frac{4\pi c R_2}{\kappa_{\rm T}}
 \cr 
 &=2.08\times10^{-3}\,(1+X)^{-1}\left(\frac{R_2}{\rsun}\right)~\msun~\rm{yr^{-1}} \,,
\end{align}
where $\kappa_{\rm T} = 0.2\,(1+X)~{\rm cm^2~g^{-1}}$ is
the Thomson scattering opacity and $X(=0.76)$ is the H-mass fraction for Pop III stars.

At the CE phase, the companion star plunges into the envelope of the donor star and spiral in.
The orbital separation after the CE phase $a_{\rm f}$ is calculated
by the energy formalism~\cite{Webbink:1984ti} which is described by
\begin{equation}
 \alpha\left(\frac{GM_{\rm{c,1}}M_2}{2a_{\rm{f}}}-\frac{GM_1M_2}{2a_{\rm{i}}}\right)
 = \frac{GM_{\rm{1}}M_{\rm{env,1}}}{\lambda R_1} \,, 
\label{eq:ce2}
\end{equation} 
where $a_{\rm i}$, $\alpha$ and $\lambda$ are the orbital separation
before the CE phase, the efficiency and the binding energy parameter, respectively.
In our standard model, we adopt $\alpha\lambda=1$.
We also calculate the $\alpha\lambda=0.01$, $0.1$ and $10$ cases in this paper.

Finally, if a binary becomes a BBH, we calculate the merger time
which is calculated from gravitational radiation reaction,
and check whether the BBH can merge within the Hubble time or not.
We repeat these calculations and take the statistics of BBH mergers.

To study the dependence of Pop III BBHs properties
on initial distribution functions and binary parameters,
we calculate 10 models with the Pop III binary population synthesis
method~\cite{Kinugawa:2014zha,Kinugawa:2015nla} in this paper.
Table~\ref{models1} shows that initial distribution functions
and the binary parameters of each model.
The columns show the model name, the IMF, the IEF,
the CE parameter $\alpha\lambda$, and the lose fraction $\beta$
of transfered stellar matter at the RLOF in each model.
 
%Table 1
\begin{table*}
\caption{The model parameters in this paper.
 Each column represents the model name, the initial mass function (IMF),
 the initial eccentricity function (IEF), 
 the common envelope (CE) parameter $\alpha\lambda$,
 and the lose fraction $\beta$ of transfer of stellar matter
 at the Roche lobe overflow (RLOF) in each model.}
\label{models1}
\begin{center}
\begin{tabular}{ccccc}
\hline
 Model & IMF & IEF & $\alpha\lambda$ & $\beta$\\
 \hline
 our standard & flat & $e$ & 1 & function \\
 IMF: logflat & $M^{-1}$ & $e$  & 1 & function \\
 IMF: Salpeter & Salpeter & $e$  & 1 & function \\
 IEF: const. & flat & const. &  1 & function \\
 IEF: $e^{-0.5}$ & flat & $e^{-0.5}$  & 1 & function \\
 $\alpha\lambda=0.01$ & flat & $e$  & 0.01 & function \\
 $\alpha\lambda=0.1$ & flat & $e$ &  0.1 & function \\
 $\alpha\lambda=10$ & flat & $e$ & 10 & function \\
 $\beta=0.5$ & flat & $e$ &  1 & 0.5 \\
 $\beta=1$ & flat & $e$ &  1 & 1 \\
 \hline
\end{tabular}
\end{center}
\end{table*}

%%%%%%%%%%%%%%%%%%%%%%%%%%%%%%%%%%%%%%%%
\section{The mass ratio distributions of binary black hole remnants}\label{sec:MR_BBHR}
%%%%%%%%%%%%%%%%%%%%%%%%%%%%%%%%%%%%%%%%

\begin{figure}[!ht]
\begin{center}
\includegraphics[width=0.5\textwidth,clip=true]{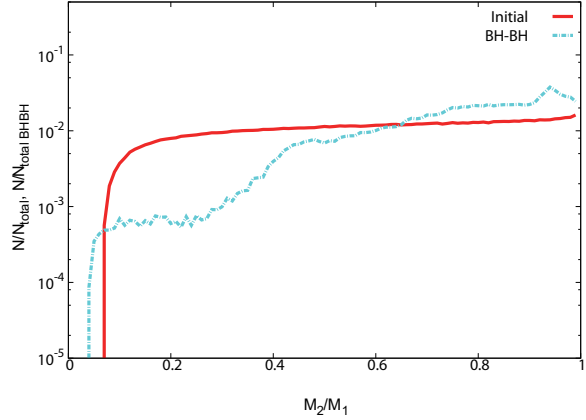}
\end{center}
 \caption{The distribution of mass ratio $M_2/M_1\leq 1$ for our standard model.
 The distributions of the initial mass ratio and
 the one when the binaries become BBHs which merge within the Hubble time,
 are shown as red and light blue lines, respectively.
 The initial mass ratio distribution is normalized by the total binary number $N_{\rm total}=10^6$,
 while the one when the binaries become merging BBHs
 is normalized by the total merging binary number $N_{\rm total \,BHBH}=128897$.}
\label{fig:massratio_standard}
\end{figure}

\begin{figure}[!ht]
\begin{center}
\includegraphics[width=0.49\textwidth,clip=true]{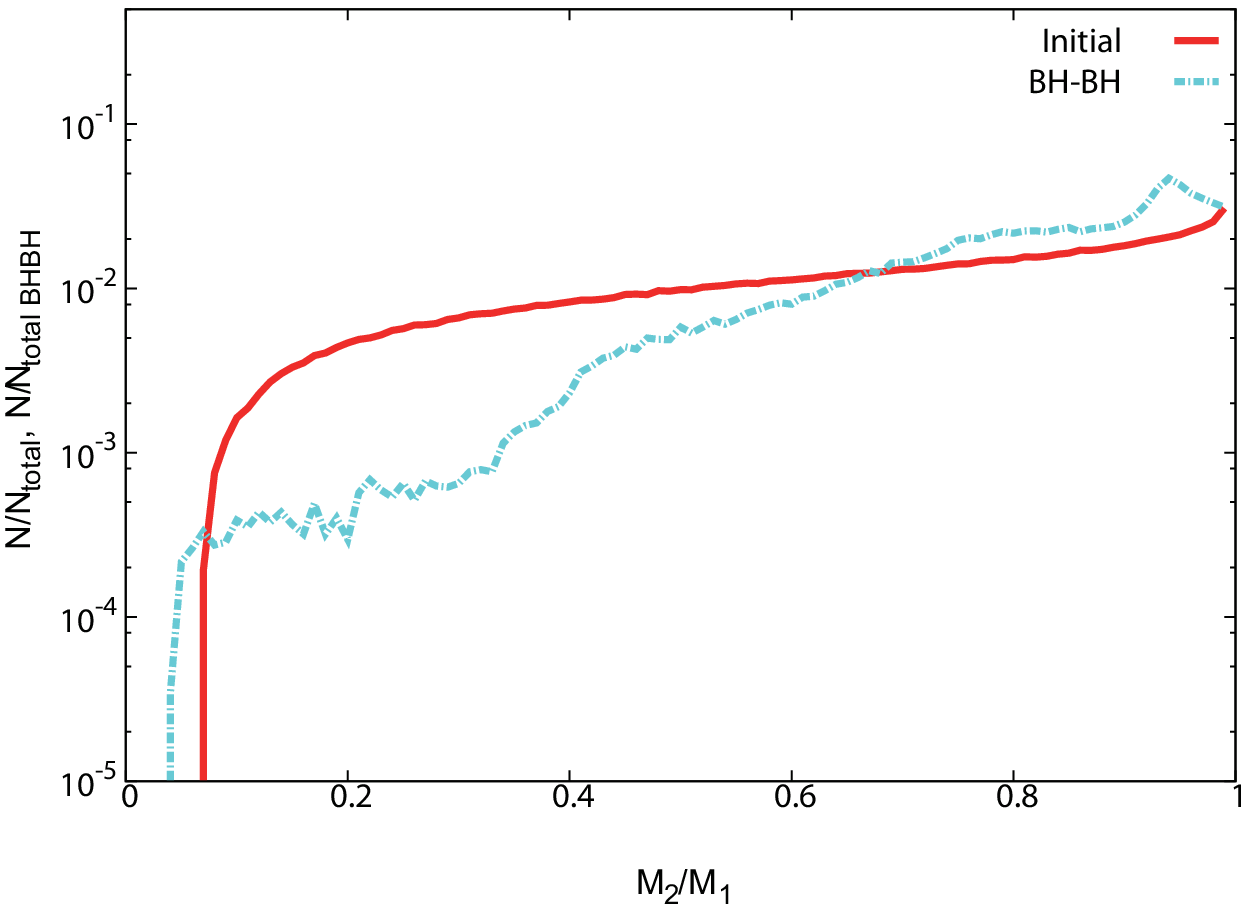}
\includegraphics[width=0.49\textwidth,clip=true]{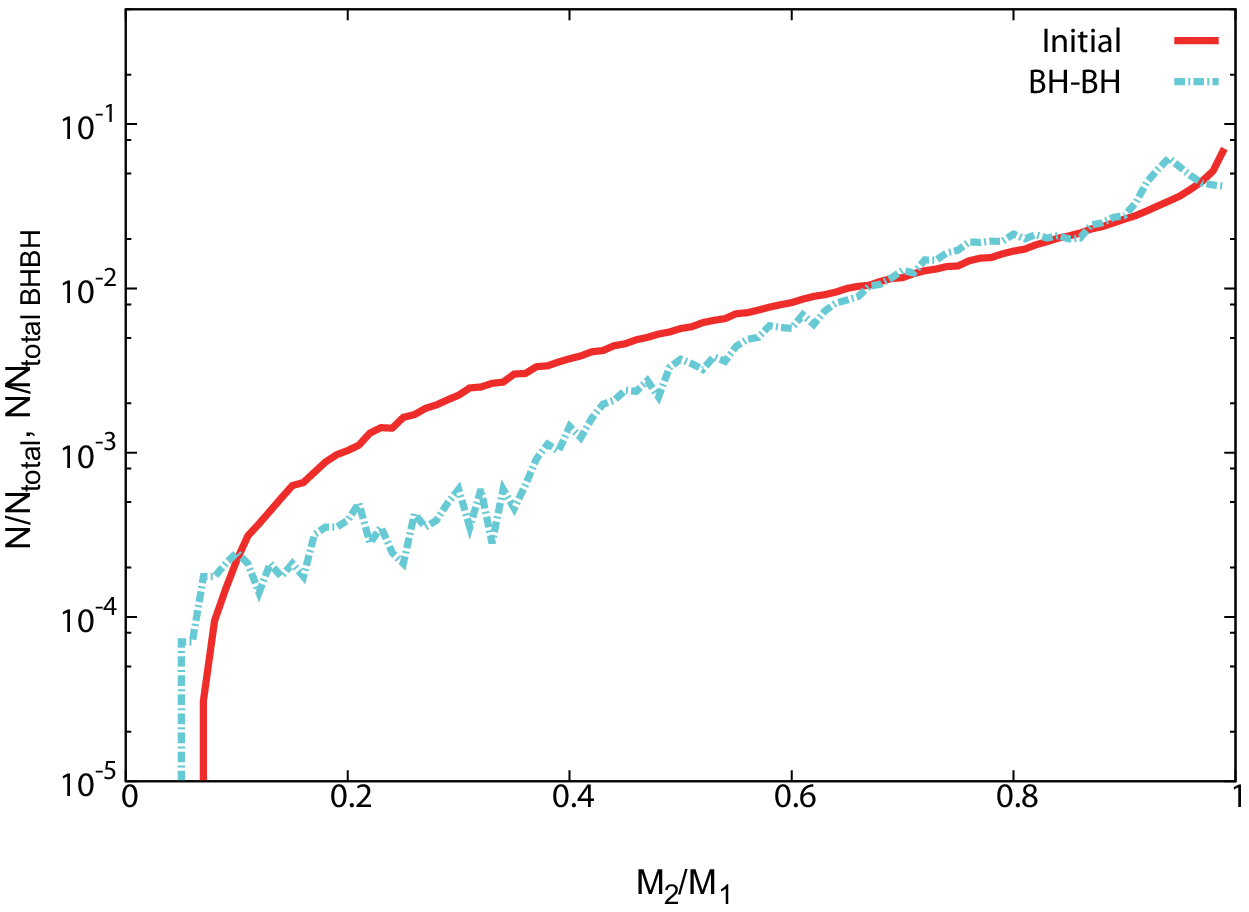}
\end{center}
 \caption{The distributions of mass ratio $M_2/M_1\leq 1$ for IMF: logflat model (left)
 and IMF: Salpeter model (right).
 The figures are same as Figure~\ref{fig:massratio_standard} but for IMF: logflat model
 and IMF: Salpeter model.}
\label{fig:massratio_logflat/salpeter}
\end{figure}

\begin{figure}[!ht]
\begin{center}
\includegraphics[width=0.49\textwidth,clip=true]{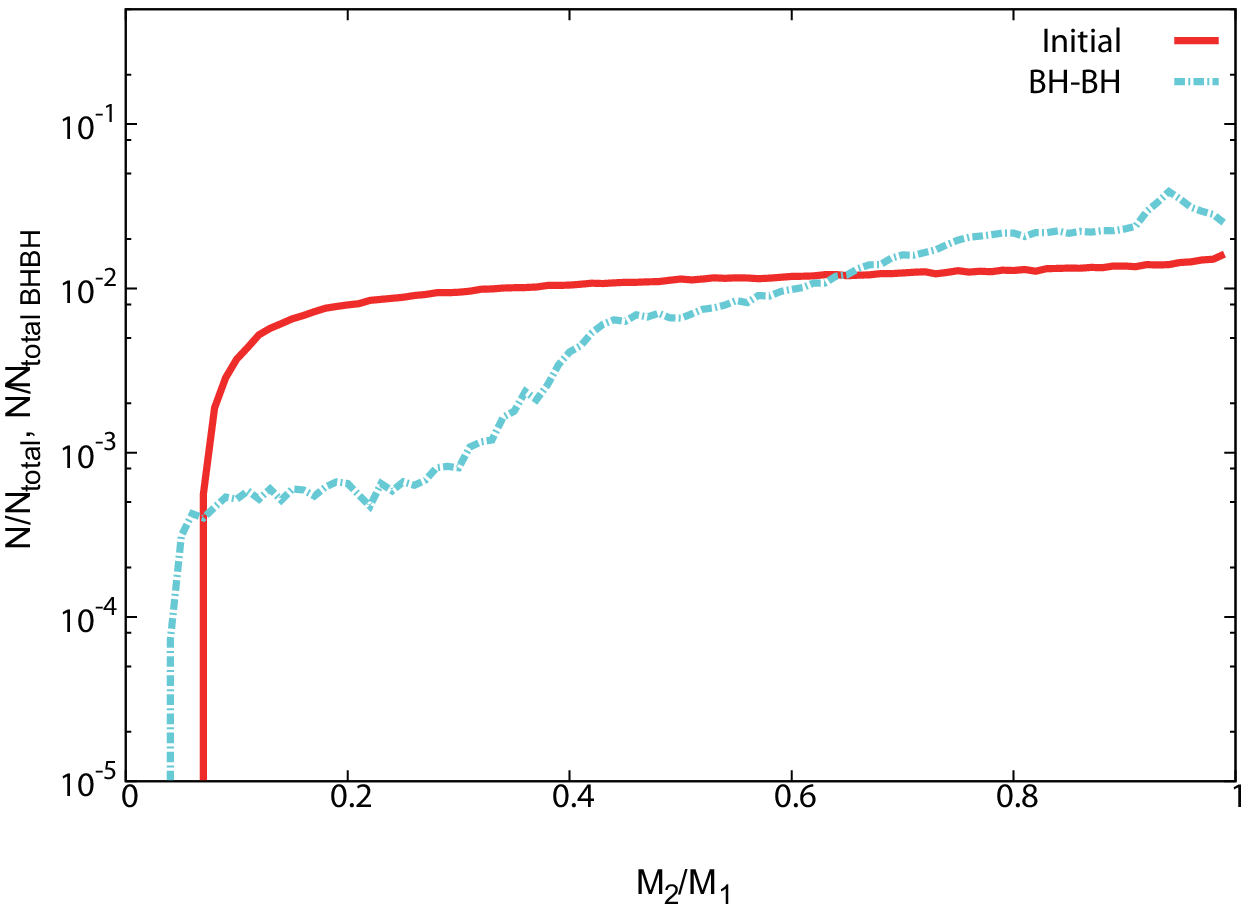}
\includegraphics[width=0.49\textwidth,clip=true]{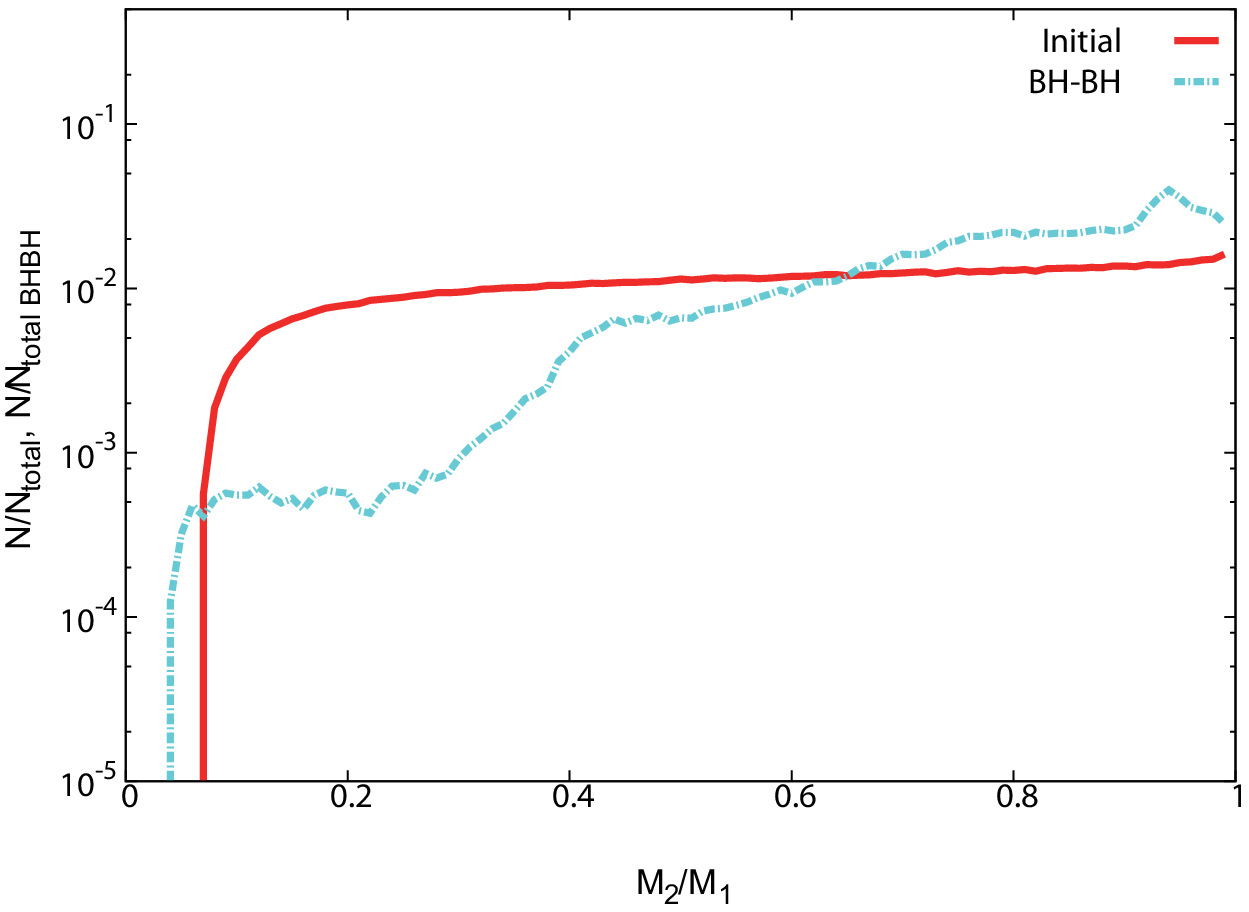}
\end{center}
 \caption{The distributions of mass ratio $M_2/M_1\leq 1$ for IEF: $e=\rm const.$ model (left)
 and IEF: $e^{-0.5}$ model (right).
 The figures are same as Figure~\ref{fig:massratio_standard} but for IEF: $e=\rm const.$ model
 and IEF: $e^{-0.5}$ model.}
\label{fig:massratio_e=const/-05}
\end{figure}

\begin{figure}[!ht]
\begin{center}
\includegraphics[width=0.49\textwidth,clip=true]{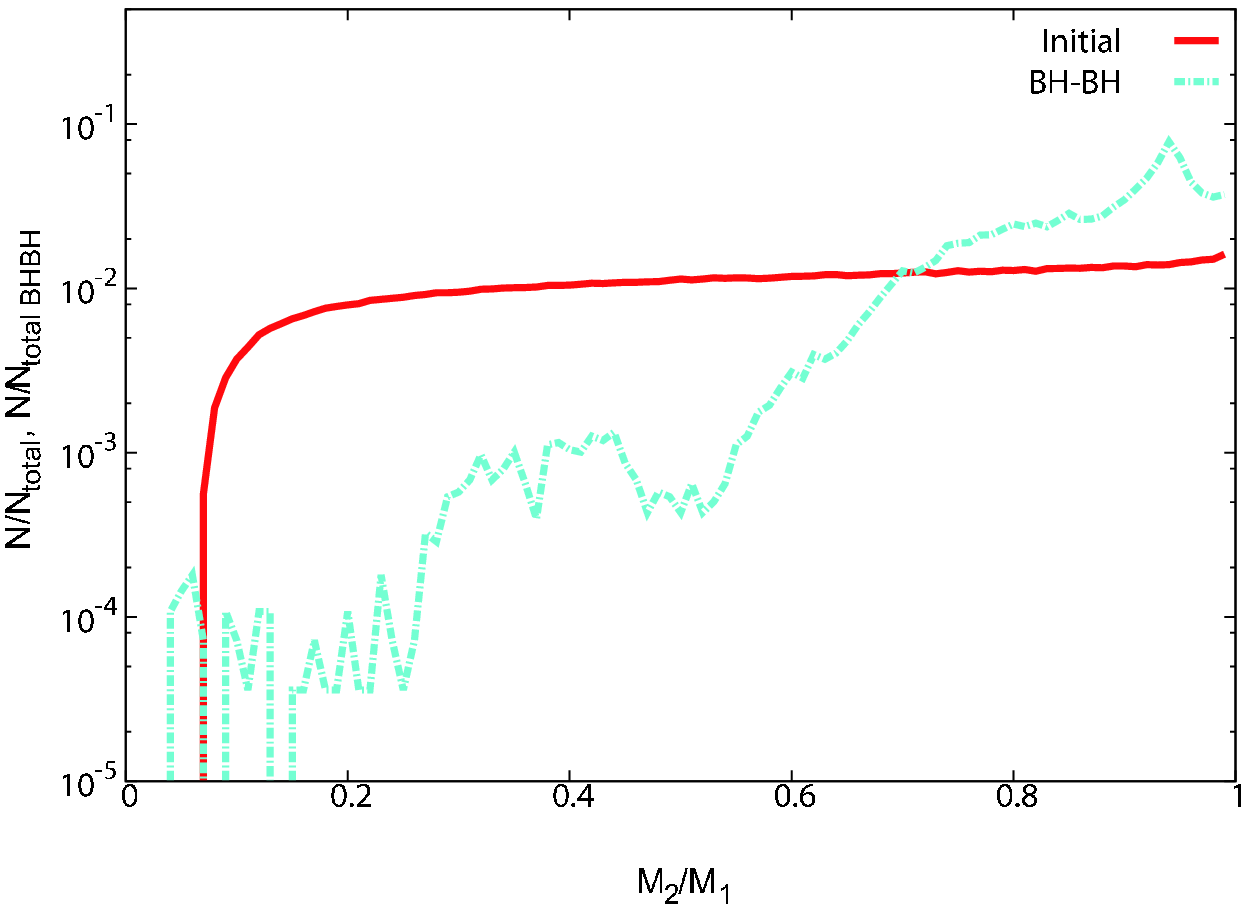}
\includegraphics[width=0.49\textwidth,clip=true]{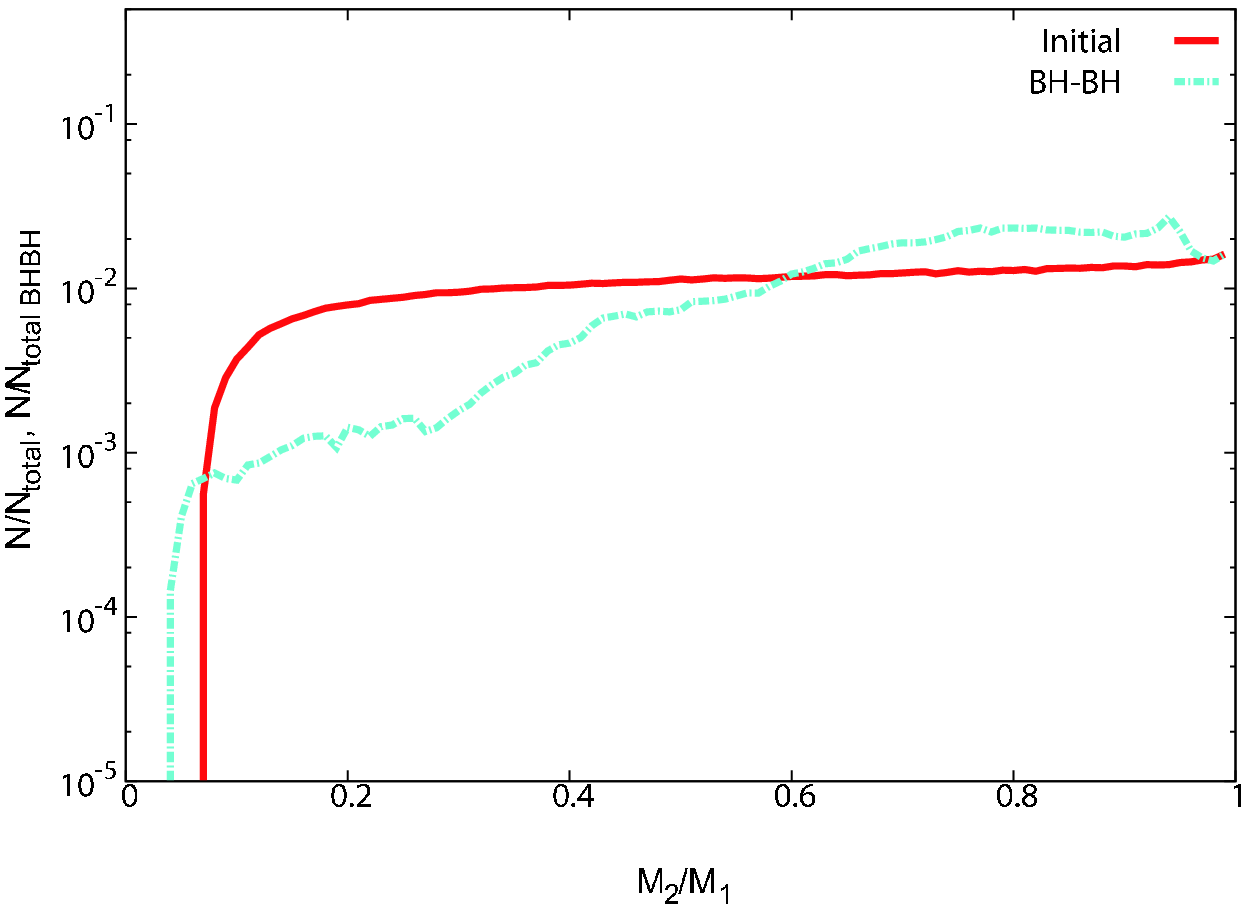}
\includegraphics[width=0.49\textwidth,clip=true]{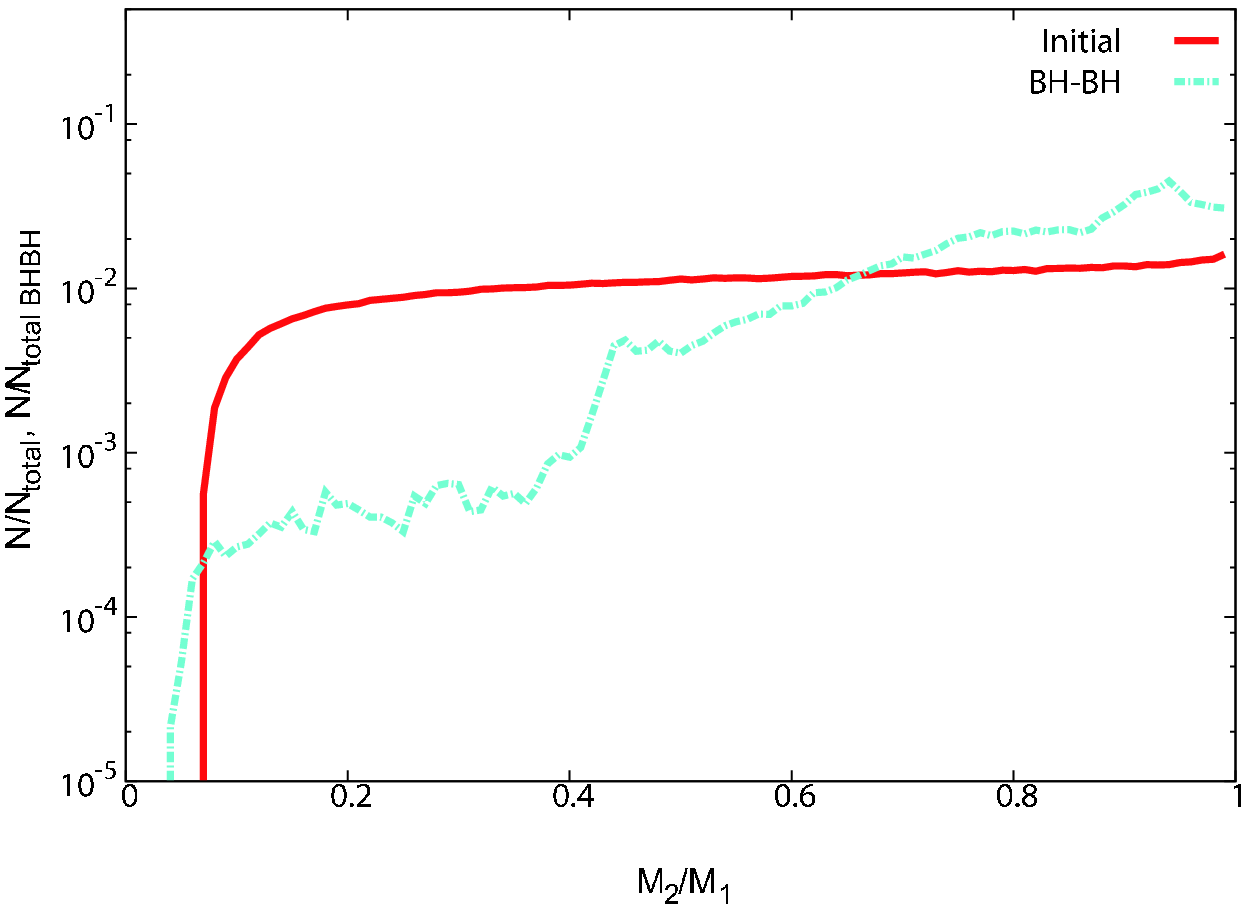}
\end{center}
 \caption{The distributions of mass ratio $M_2/M_1\leq 1$ for $\alpha\lambda=0.01$ model (top-left),
 $\alpha\lambda=0.1$ model (top-right) and $\alpha\lambda=10$ model (bottom).
 The figures are same as Figure \ref{fig:massratio_standard} but for $\alpha\lambda=0.01$ model,
 $\alpha\lambda=0.1$ model and $\alpha\lambda=10$ model.}
\label{fig:massratio_al=001/01/10}
\end{figure}

\begin{figure}[!ht]
\begin{center}
\includegraphics[width=0.49\textwidth,clip=true]{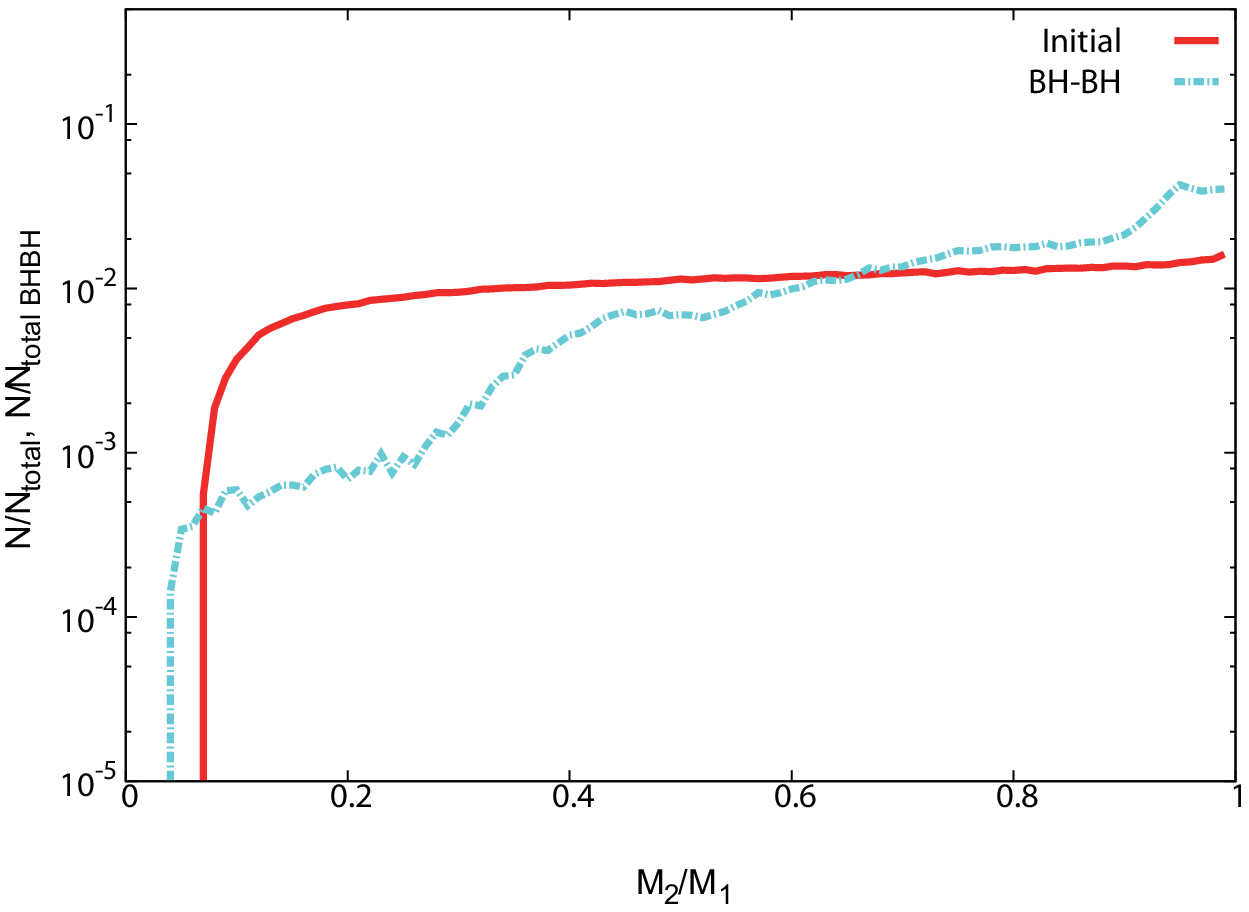}
\includegraphics[width=0.49\textwidth,clip=true]{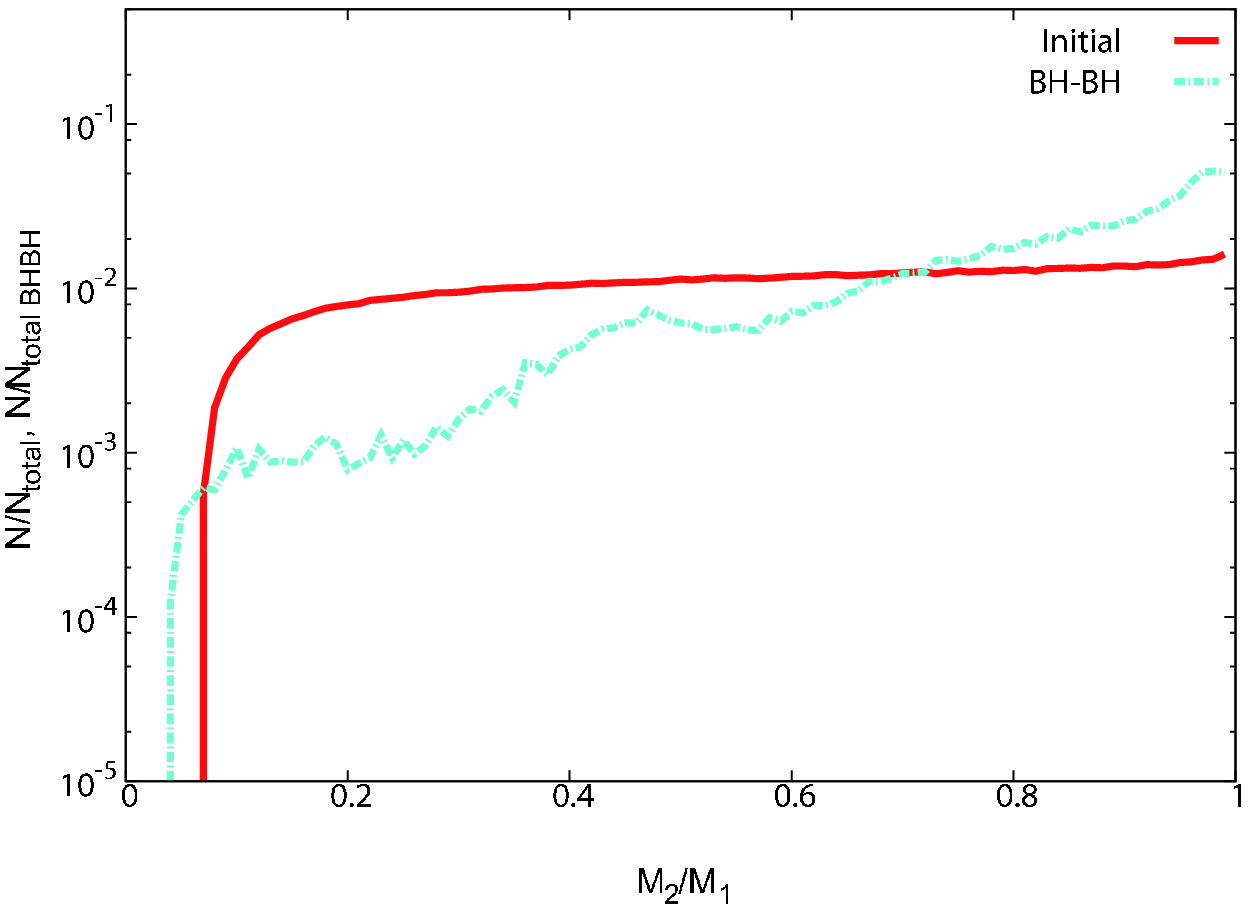}
\end{center}
 \caption{The distributions of mass ratio $M_2/M_1\leq 1$ for $\beta=0.5$ model (left)
 and $\beta=1$ model (right).
 The figures are same as Figure~\ref{fig:massratio_standard} but for $\beta=0.5$ model
 and $\beta=1$ model.}
\label{fig:massratio_b=05/1}
\end{figure}

Figures~\ref{fig:massratio_standard}--\ref{fig:massratio_b=05/1} show
the initial mass ratio distributions and the mass ratio distributions of merging BBHs.
The RLOF tends to make binaries to be equal mass.
Thus, the BBHs mass ratio distributions depend on
how many binaries evolve via the RLOF.
The Pop III stars with mass $< 50~\msun$ evolve as a blue giant.
Thus, in the case of the IMF that light stars are majority,
the binaries tend to evolve only via the RLOF, not via the CE phase. 
Therefore, the steeper IMFs tend to derive many equal mass BBHs.
In this calculation, since we adopt the minimum mass ratio as $10\msun/M_1$,
the initial mass ratio distribution of models
with the IMF that light stars are majority, is upward to $M_2/M_1=1$ ab initio
(see Figs.~\ref{fig:massratio_standard} and \ref{fig:massratio_logflat/salpeter}).

%%%%%%%%%%%%%%%%%%%%%%%%%%%%%%%%%%%%%%%%%%%%%%%%%%
%%%%%%%%%%%%%%%%%%%%%%%%%%%%%%%%%%%%%%%%%%%%%%%%%%IMF
%%%%%%%%%%%%%%%%%%%%%%%%%%%%%%%%%%%%%%%%%%%%%%%%%%

On the other hand, if we change the IEF, the mass ratio distribution
does not change much. Thus, the dependence of  IEF is not so large
(see Figs.~\ref{fig:massratio_standard} and \ref{fig:massratio_e=const/-05}).

%%%%%%%%%%%%%%%%%%%%%%%%%%%%%%%%%%%%%%%%%%%%%%%%%%IEF
%%%%%%%%%%%%%%%%%%%%%%%%%%%%%%%%%%%%%%%%%%%%%%%%%%

As for the CE parameter dependence
(see Figs.~\ref{fig:massratio_standard} and \ref{fig:massratio_al=001/01/10}),
small mass ratio binaries in the model of $\alpha\lambda=0.001$ 
are much fewer than those in the other models.
In the $\alpha\lambda=0.001$ model, all the binaries which evolve via the CE phase,
merge during the CE phase due to too small $\alpha\lambda$.
Thus, the merging BBHs in this model evolve only via the RLOF,
and become equal mass by the RLOF.
There is not large change between the models
with the CE parameters $\alpha\lambda=0.1$ and $10$.

%%%%%%%%%%%%%%%%%%%%%%%%%%%%%%%%%%%%%%%%%%%%%%%%%%CE parameter
%%%%%%%%%%%%%%%%%%%%%%%%%%%%%%%%%%%%%%%%%%%%%%%%%%

As for the mass loss fraction $\beta$
(see Figs.~\ref{fig:massratio_standard} and \ref{fig:massratio_b=05/1}),
when $\beta$ becomes large, there are three effects.
First, binaries tend not to become a CE phase.
Second, the mass accretion by a RLOF becomes not to be effective
Third, a RLOF tends to finish early. 
The first effect make binaries evolve via a RLOF.
However, the second and third effects have a negative impact on the tendency
to become equal mass.
Thus, the mass ratio distributions of $\beta=0.5$ and $1$ models
look like less change from that of our standard model.

%%%%%%%%%%%%%%%%%%%%%%%%%%%%%%%%%%%%%%%%%%%%%%%%%%beta

%%%%%%%%%%%%%%%%%%%%%%%%%%%%%%%%%%%%%%%%
\section{The spin distributions of binary black hole remnants}\label{S_BBHR}
%%%%%%%%%%%%%%%%%%%%%%%%%%%%%%%%%%%%%%%%

When the Pop III star becomes a BH, we calculate the BH's spin.
If the estimated spin of the BH is more than the Thorne limit
$q_{\rm Thorne}=0.998$~\cite{Thorne:1974ve},
we assign the non-dimensional spin parameter $q=q_{\rm Thorne}$.
 
We ignore the spin up by the accretion during a mass transfer
after the star became a BH. The reason is described as follows.
The spin up by the accretion is calculated as
\begin{align}
 \delta q = & \frac{\delta J}{GM^2_{\rm BH}/c}
 \cr
 = & \sqrt{12}\frac{\delta M}{M_{\rm BH}} \,,
\end{align}
where $\delta J$, $M_{\rm BH}$, $\delta M$ are the gain of angular momentum,
the BH's mass, and the gain of BH's mass, respectively.
Since the accretion rate of the BH during a RLOF is the Eddington rate,
the gain of BH's mass is 
\begin{equation}
\delta M \sim \dot{M}_{\rm Edd}\,t_{\rm life} \,,
\end{equation}
where $t_{\rm life}$ is the lifetime of the Pop III star.
The Eddington accretion rate is given by
\begin{equation}
 \dot{M}_{\rm Edd} \sim 10^{-7}~\left(\frac{M_{\rm BH}}{30~\msun}\right)~\msun~{\rm yr^{-1}} \,,
\end{equation}
and the lifetime of massive star is $t_{\rm life}\sim1~{\rm Myr}$.
As the result, we have $\delta q \sim 0.01$, and the spin up by the accretion during a RLOF is negligible.
On the other hand, in the case of a CE phase, the accretion rate during the CE phase is
$\dot{M}\sim10^{-3}~\msun~\rm yr^{-1}$~\cite{Ivanova:2012vx},
and the timescale of the CE phase is about a thermal timescale of red giant
$t_{\rm KH}\sim 10^2~\rm yr$ or less than the thermal timescale.
As the result, we have $\delta q \leq 0.1$, and the spin up by the accretion
during a CE phase is negligible, too.

\begin{figure}[!ht]
\begin{center}
\includegraphics[width=0.5\textwidth,clip=true]{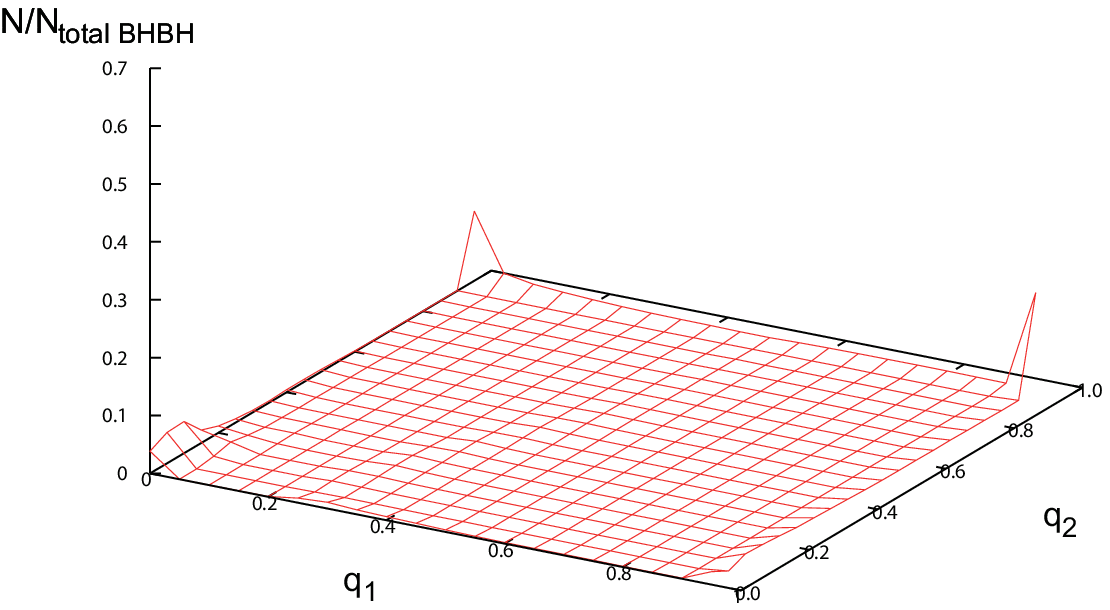}
        \subfloat[$0<q_1<0.05$]{
            \includegraphics[scale=0.4]{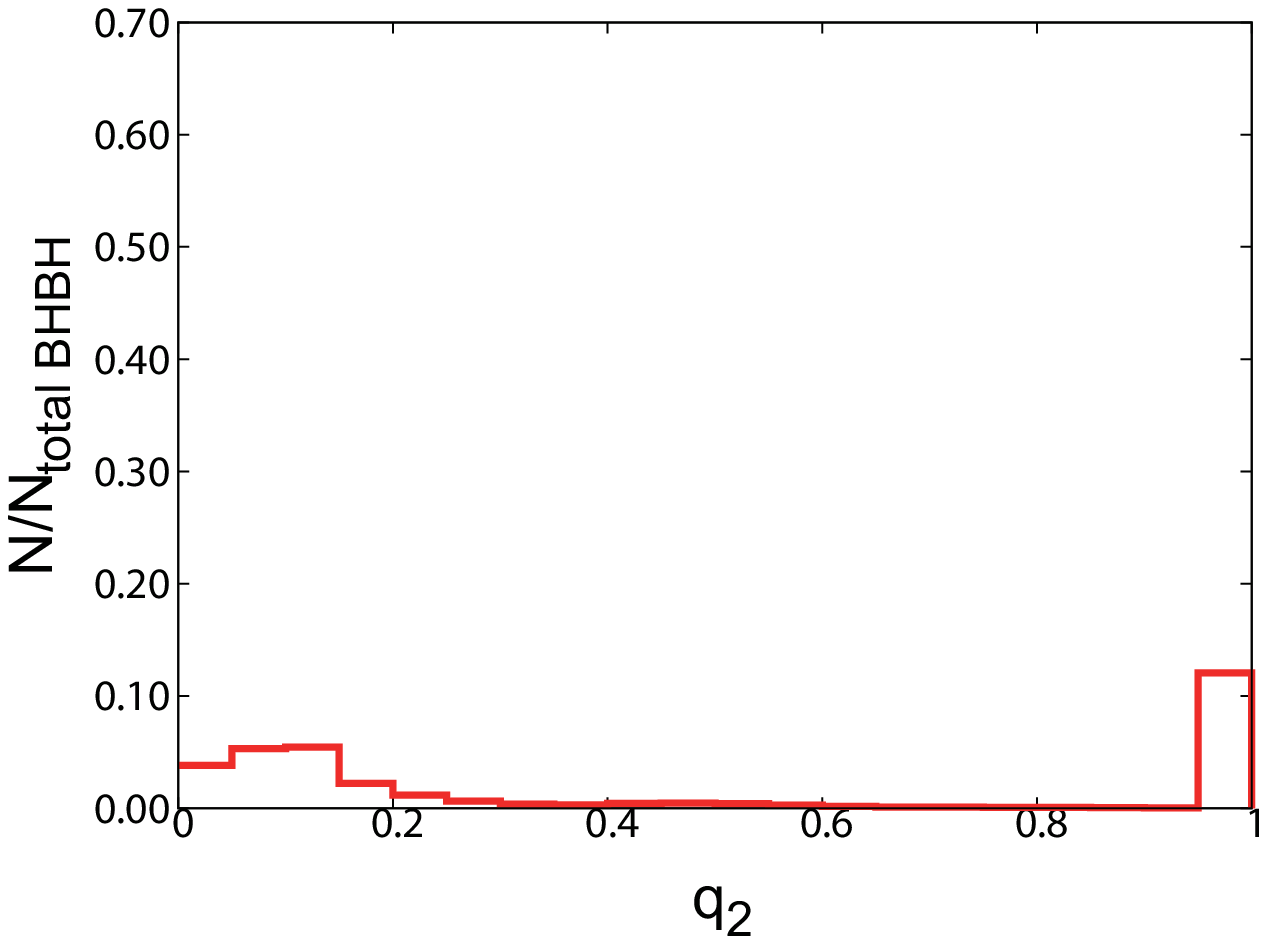}
        }
        \subfloat[$0.95<q_1<0.998$]{
            \includegraphics[scale=0.4]{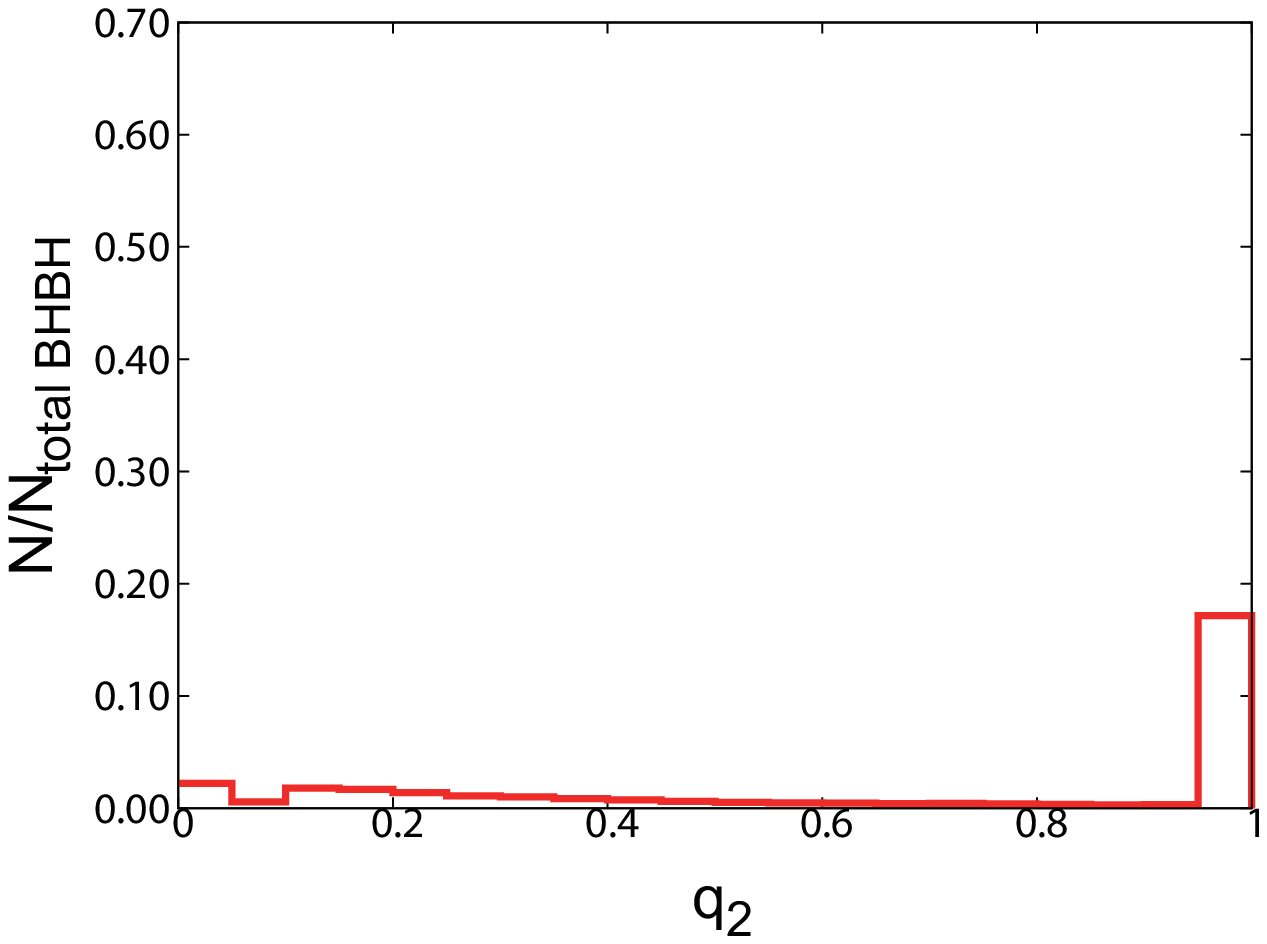}
        }
\end{center}
 \caption{(Top) The distribution of spin parameters for our standard model.
 This figure shows the  distribution of spin parameters
 when each star becomes BH.
 $q_1$ and $q_2$ are the spin parameters of the primary and the secondary BHs, respectively.
 This distribution when the binaries become merging BBHs is normalized
 by the total merging BBH number $N_{\rm total \,BHBH}=128897$
 with the grid separation being $\Delta q_1 = \Delta q_2=0.05$.
 (Bottom) Cross section views of distribution of spin parameter for our standard model.
 (a) The distribution of $q_2$ for $0 < q_1 < 0.05$.  We can see  that $q_2$ distribution
 has bimodial peaks at  $0<q_2<0.15$ and $0.95<q_2<0.998$.
 (b) The distribution of $q_2$ for $0.95 < q_1 < 0.998$. We see that the large value of $q_2$
 is the majority
 so that  there is a group in which both $q_1$ and $q_2$ are large.
}
 \label{fig:spin1spin2standard}
\end{figure}

\begin{figure}[!ht]
\begin{center}
\includegraphics[width=0.5\textwidth,clip=true]{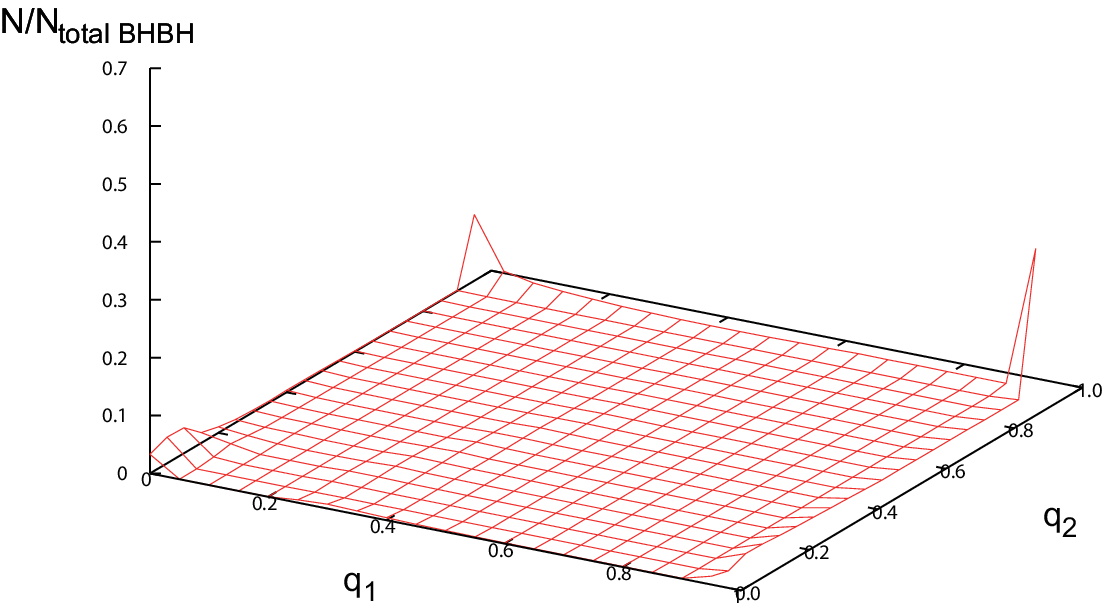}
        \subfloat[$0<q_1<0.05$]{
            \includegraphics[scale=0.4]{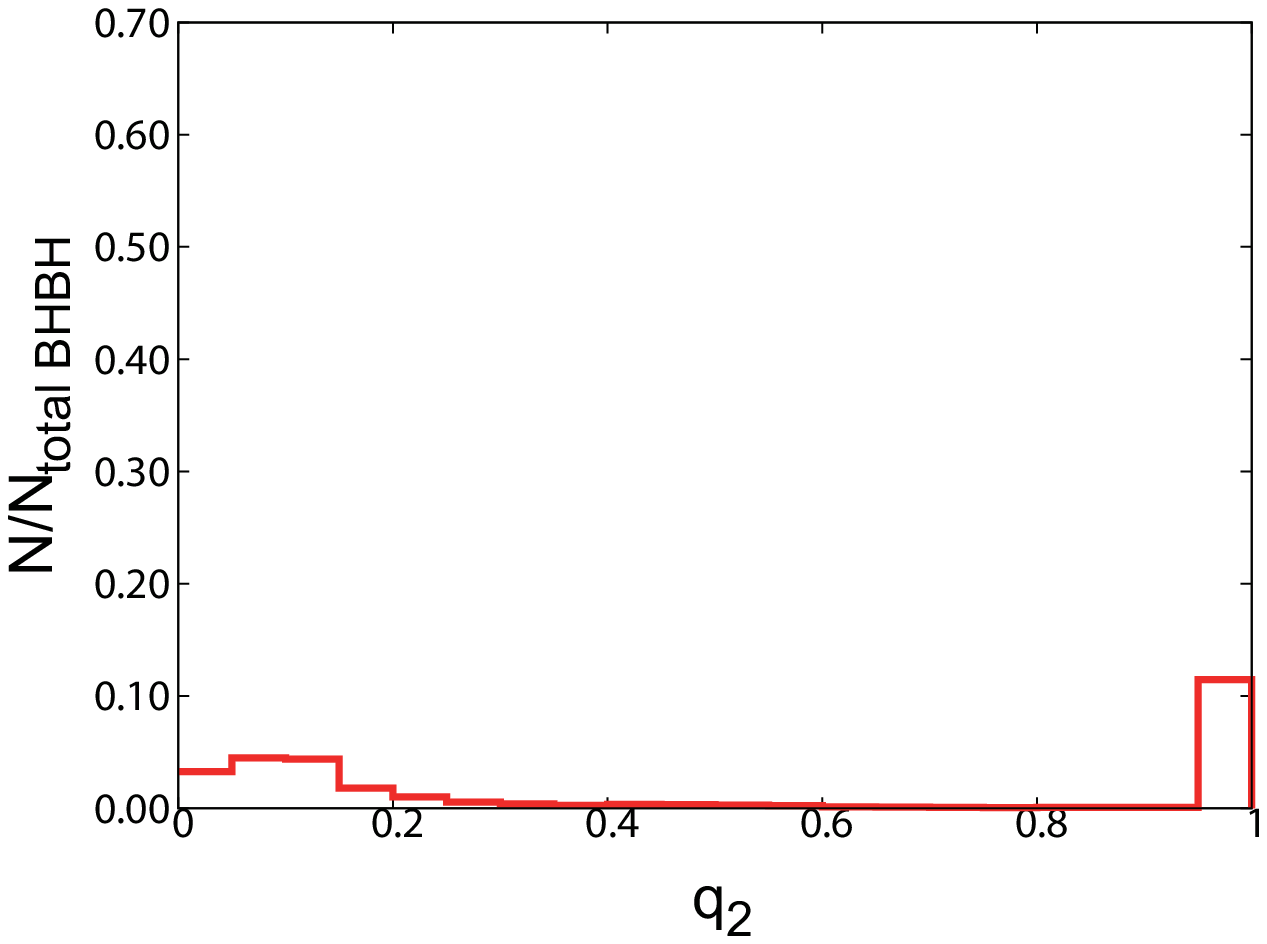}
        }
        \subfloat[$0.95<q_1<0.998$]{
            \includegraphics[scale=0.4]{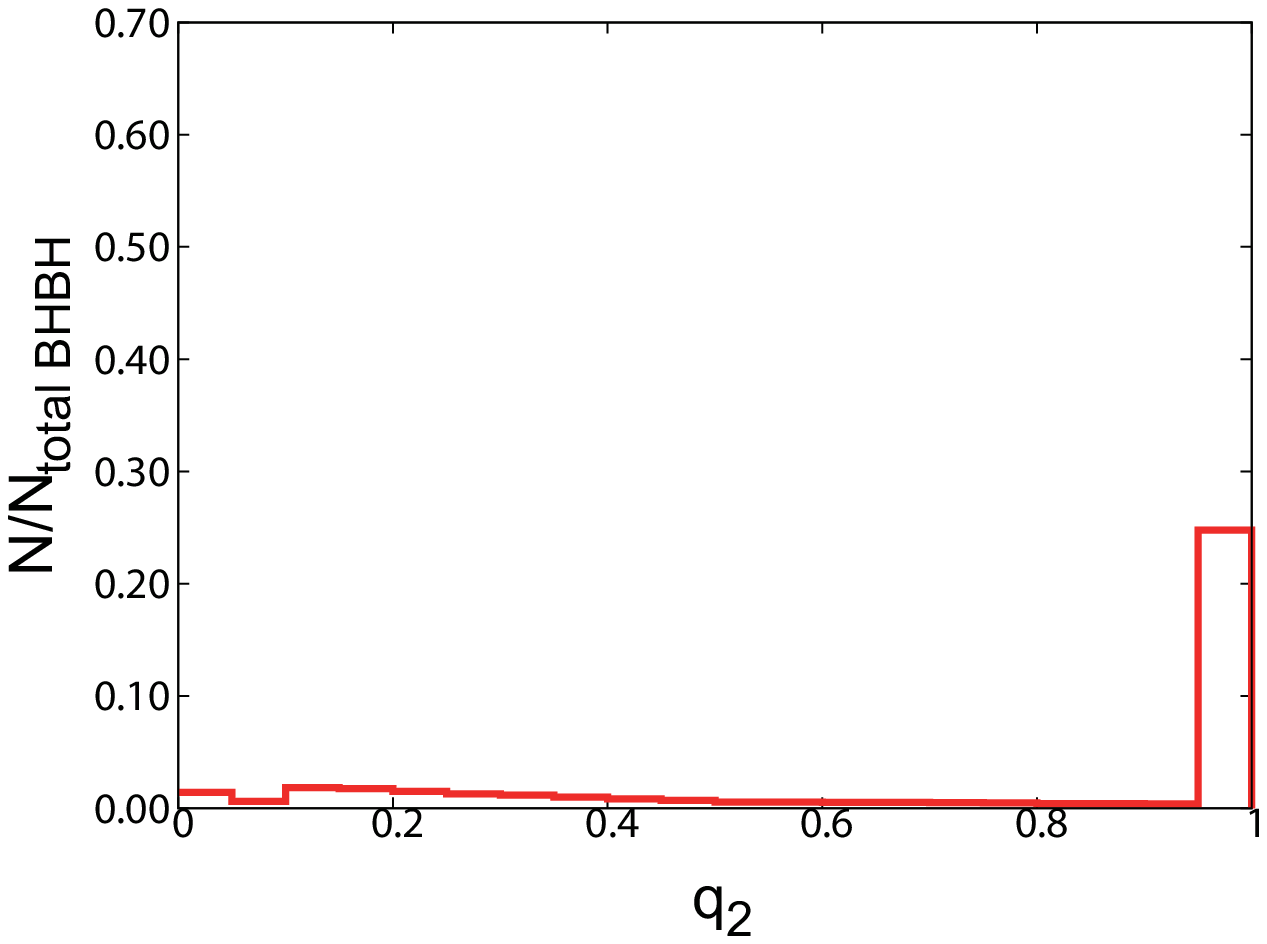}
        }
\end{center}
 \caption{(Top) The distribution of spin parameters for IMF: logflat model.
 $N_{\rm total \,BHBH}=87596$.
 (Bottom) Cross section views of distribution of spin parameter for IMF: logflat model.
 These figures are same as Figure~\ref{fig:spin1spin2standard} but for IMF: logflat model.}
 \label{fig:spin1spin2logflat}
\end{figure}

\begin{figure}[!ht]
\begin{center}
\includegraphics[width=0.5\textwidth,clip=true]{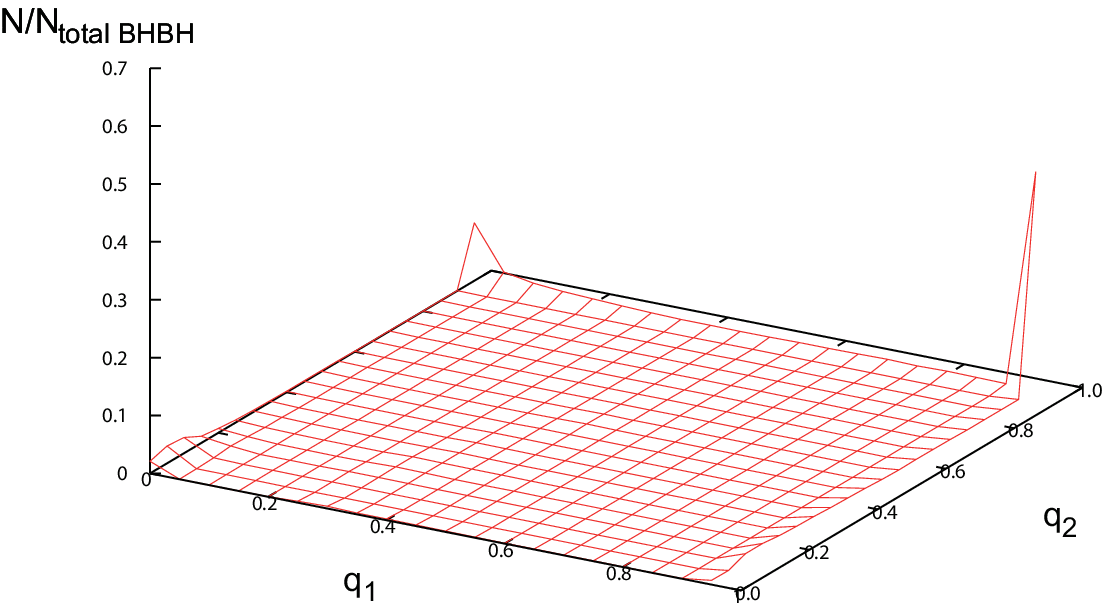}
        \subfloat[$0<q_1<0.05$]{
            \includegraphics[scale=0.4]{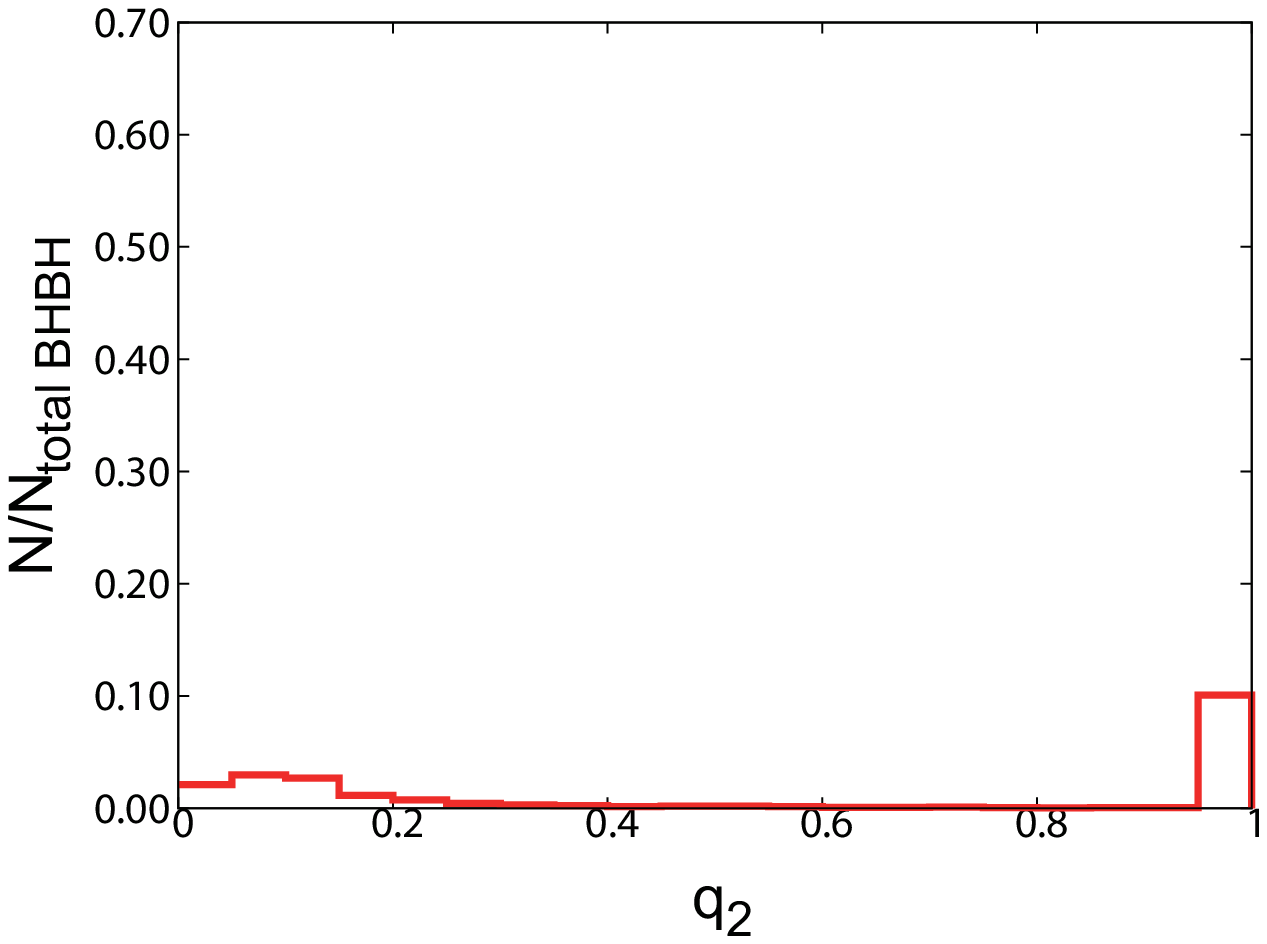}
        }
        \subfloat[$0.95<q_1<0.998$]{
            \includegraphics[scale=0.4]{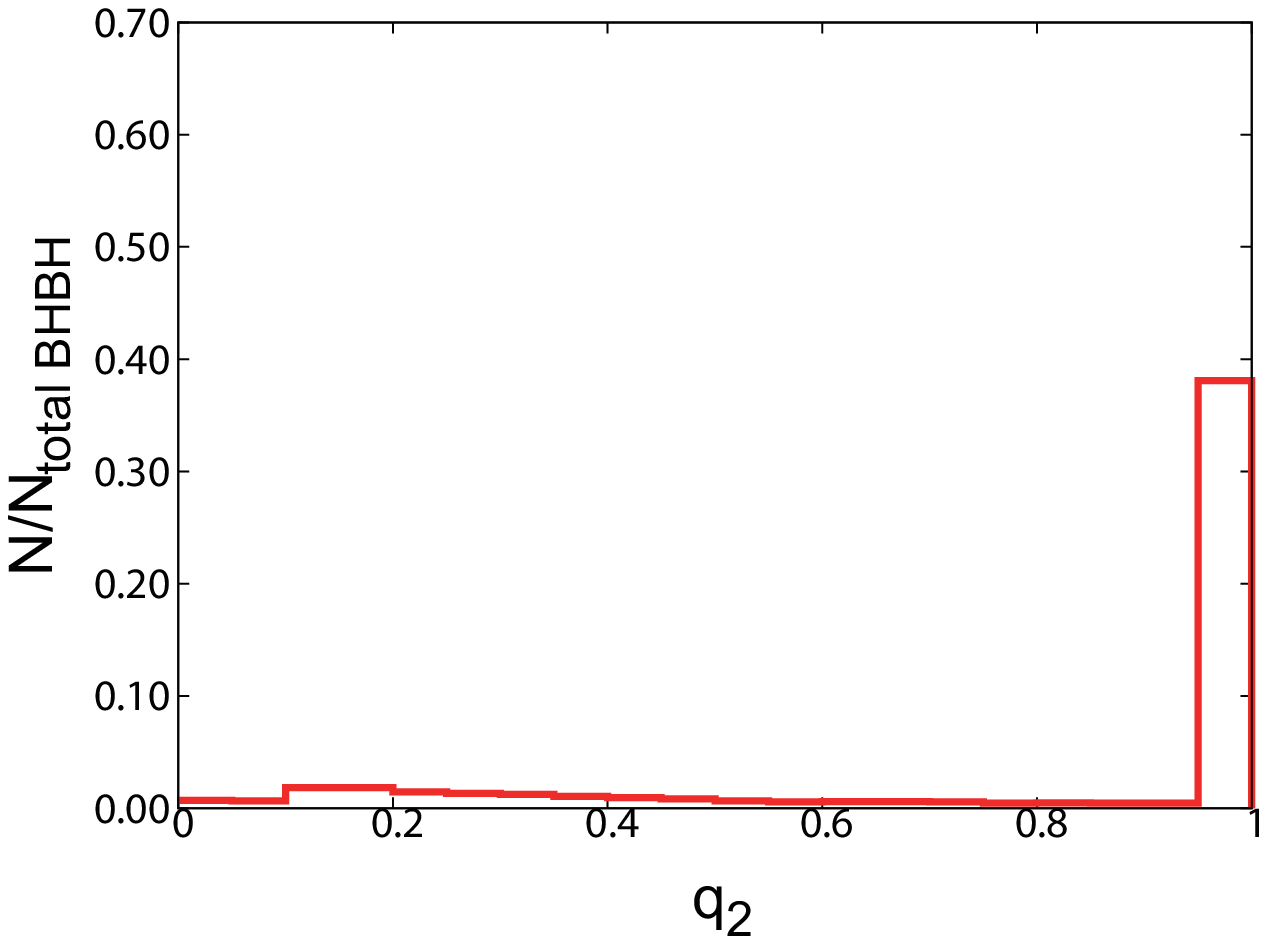}
        }
\end{center}
 \caption{(Top) The distribution of spin parameters for IMF: Salpeter model. 
 $N_{\rm total \,BHBH}=28376$.
 (Bottom) Cross section views of distribution of spin parameter for IMF: Salpeter model.
 This figure is same as Figure~\ref{fig:spin1spin2standard} but for IMF: Salpeter model.}
 \label{fig:spin1spin2salpeter}
\end{figure}

\begin{figure}[!ht]
\begin{center}
\includegraphics[width=0.5\textwidth,clip=true]{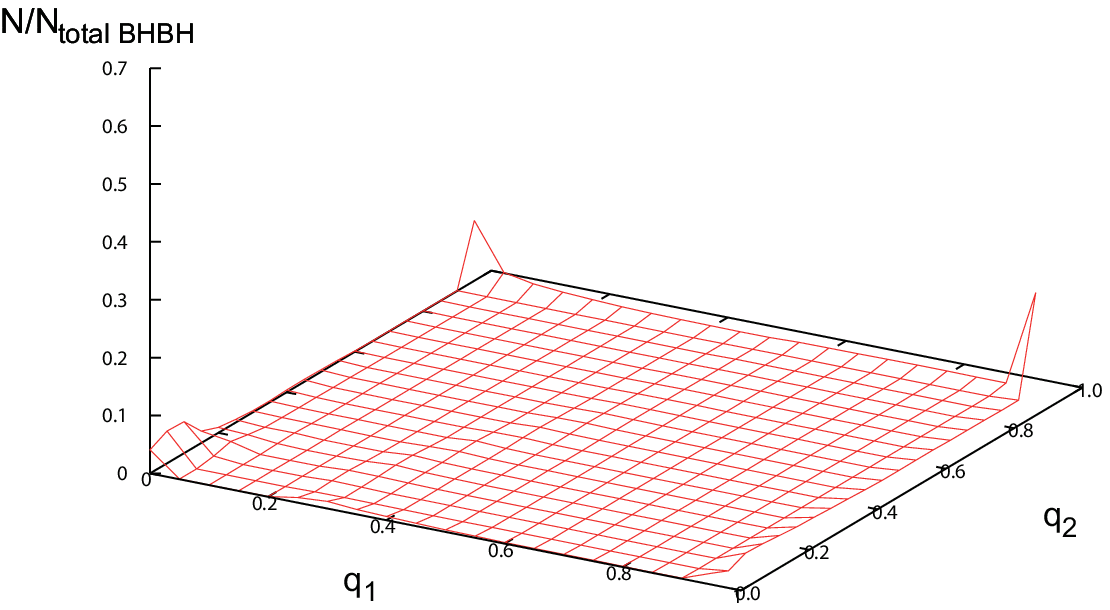}
        \subfloat[$0<q_1<0.05$]{
            \includegraphics[scale=0.4]{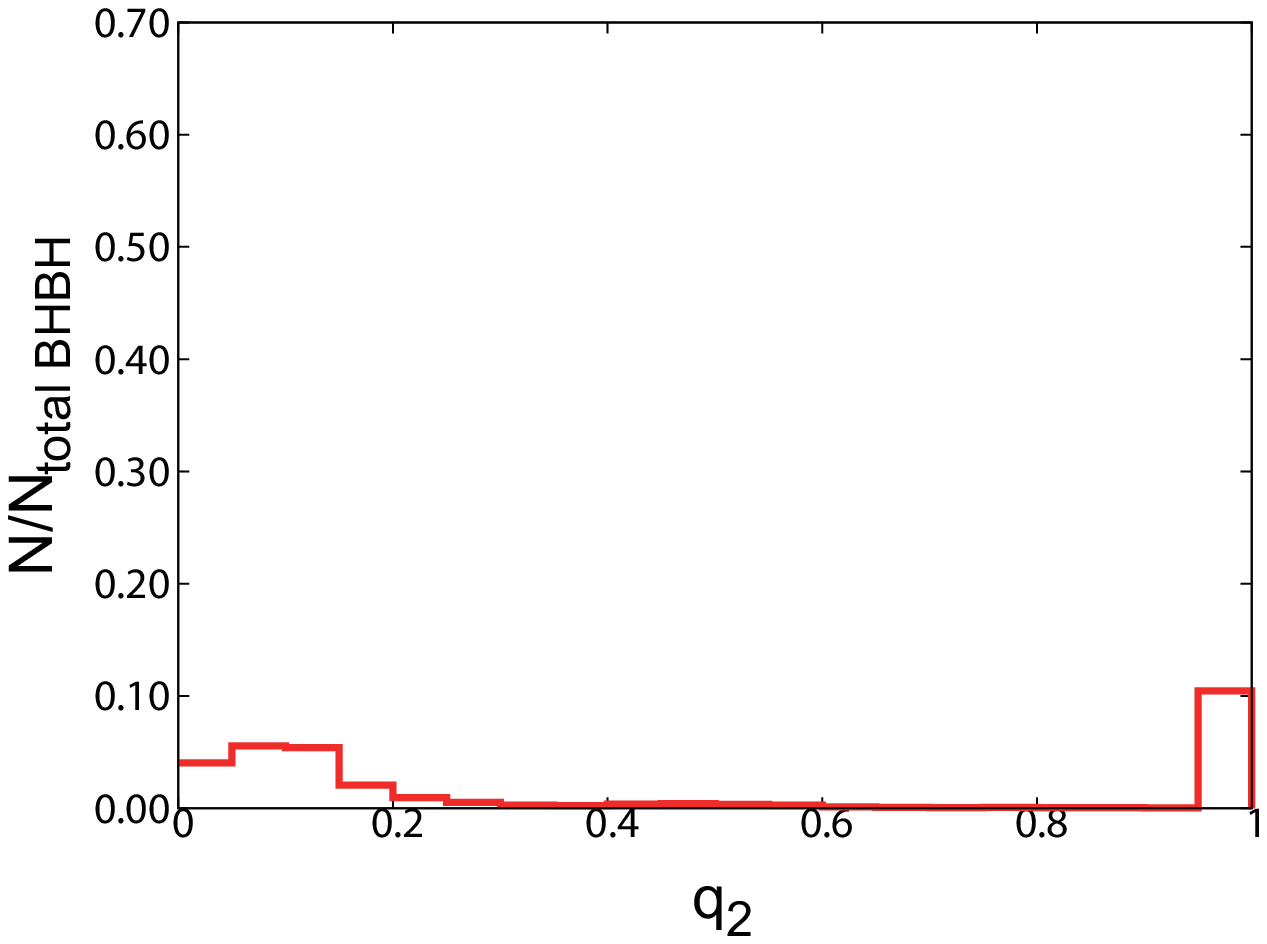}
        }
        \subfloat[$0.95<q_1<0.998$]{
            \includegraphics[scale=0.4]{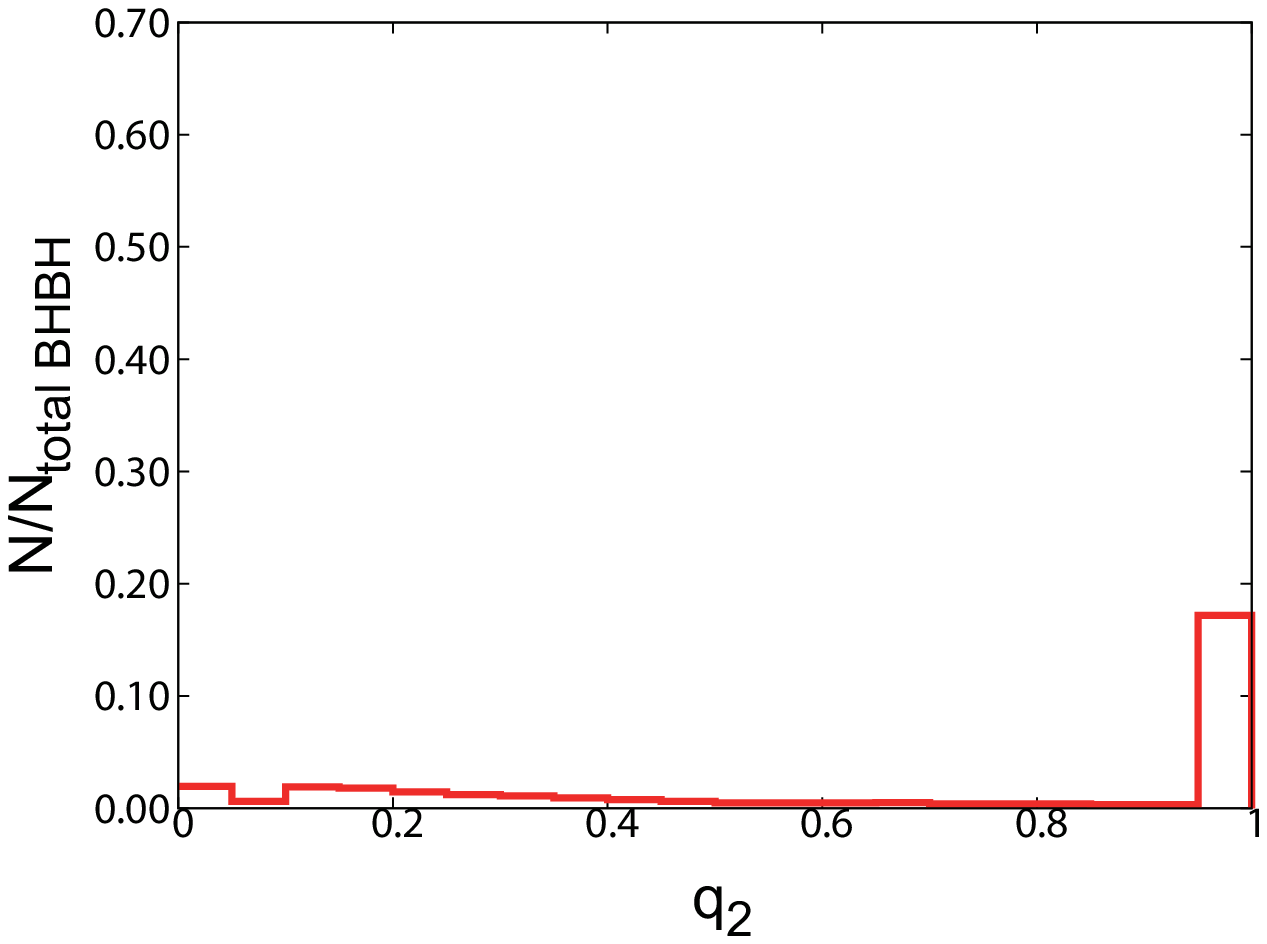}
        }
\end{center}
 \caption{(Top) The distribution of spin parameters for IEF: $e=\rm const.$ model.
 $N_{\rm total \,BHBH}=124711$.
 (Bottom) Cross section views of distribution of spin parameter for IEF: $e=\rm const.$ model.
 This figure is same as Figure~\ref{fig:spin1spin2standard} but for IEF: $e=\rm const.$ model.}
 \label{fig:spin1spin2e=const}
\end{figure}

\begin{figure}[!ht]
\begin{center}
\includegraphics[width=0.5\textwidth,clip=true]{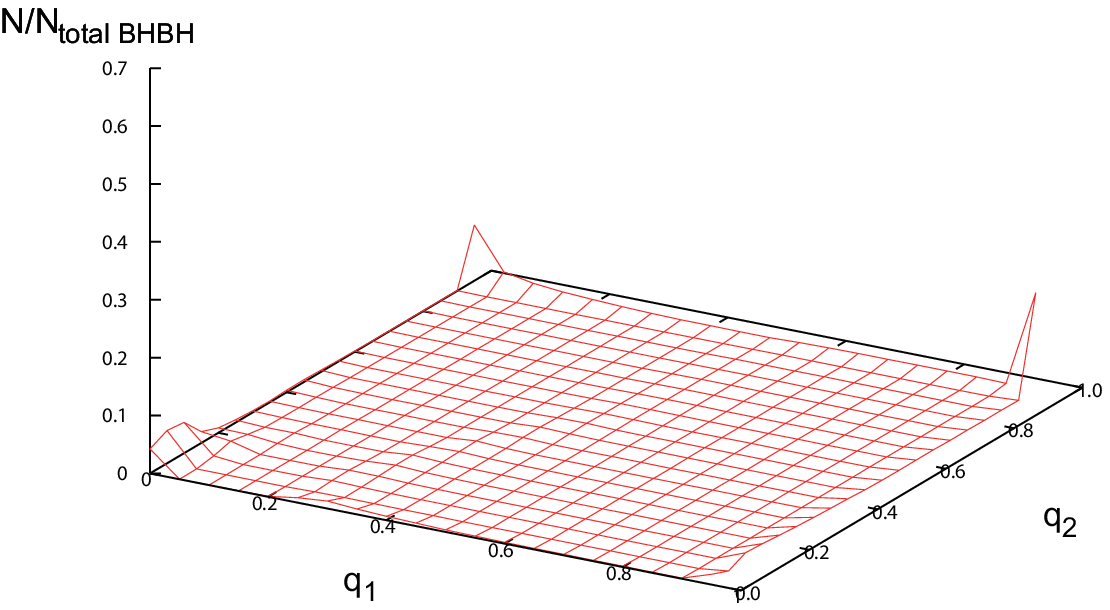}
        \subfloat[$0<q_1<0.05$]{
            \includegraphics[scale=0.4]{spin1spin2q0e=const.eps}
        }
        \subfloat[$0.95<q_1<0.998$]{
            \includegraphics[scale=0.4]{spin1spin2q1e=const.eps}
        }
\end{center}
 \caption{(Top) The distribution of spin parameters for IEF: $e^{-0.5}$ model.
 $N_{\rm total \,BHBH}=121495$.
 (Bottom) Cross section views of distribution of spin parameter for IEF: $e^{-0.5}$ model.
 This figure is same as Figure~\ref{fig:spin1spin2standard} but for IEF: $e^{-0.5}$ model.}
 \label{fig:spin1spin2e-05}
\end{figure}

\begin{figure}[!ht]
\begin{center}
\includegraphics[width=0.5\textwidth,clip=true]{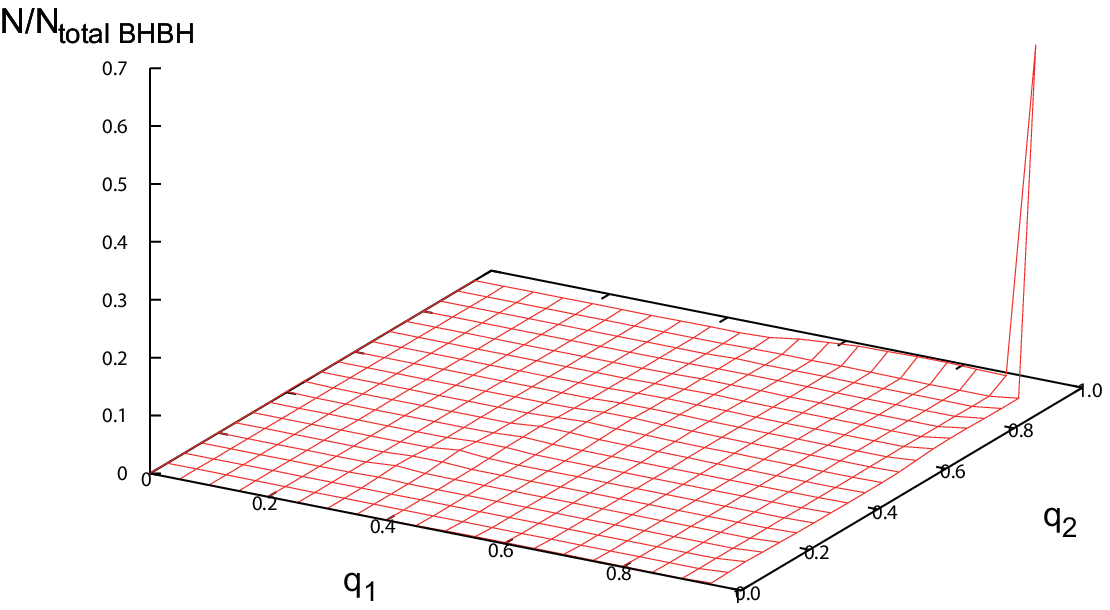}
        \subfloat[$0<q_1<0.05$]{
            \includegraphics[scale=0.4]{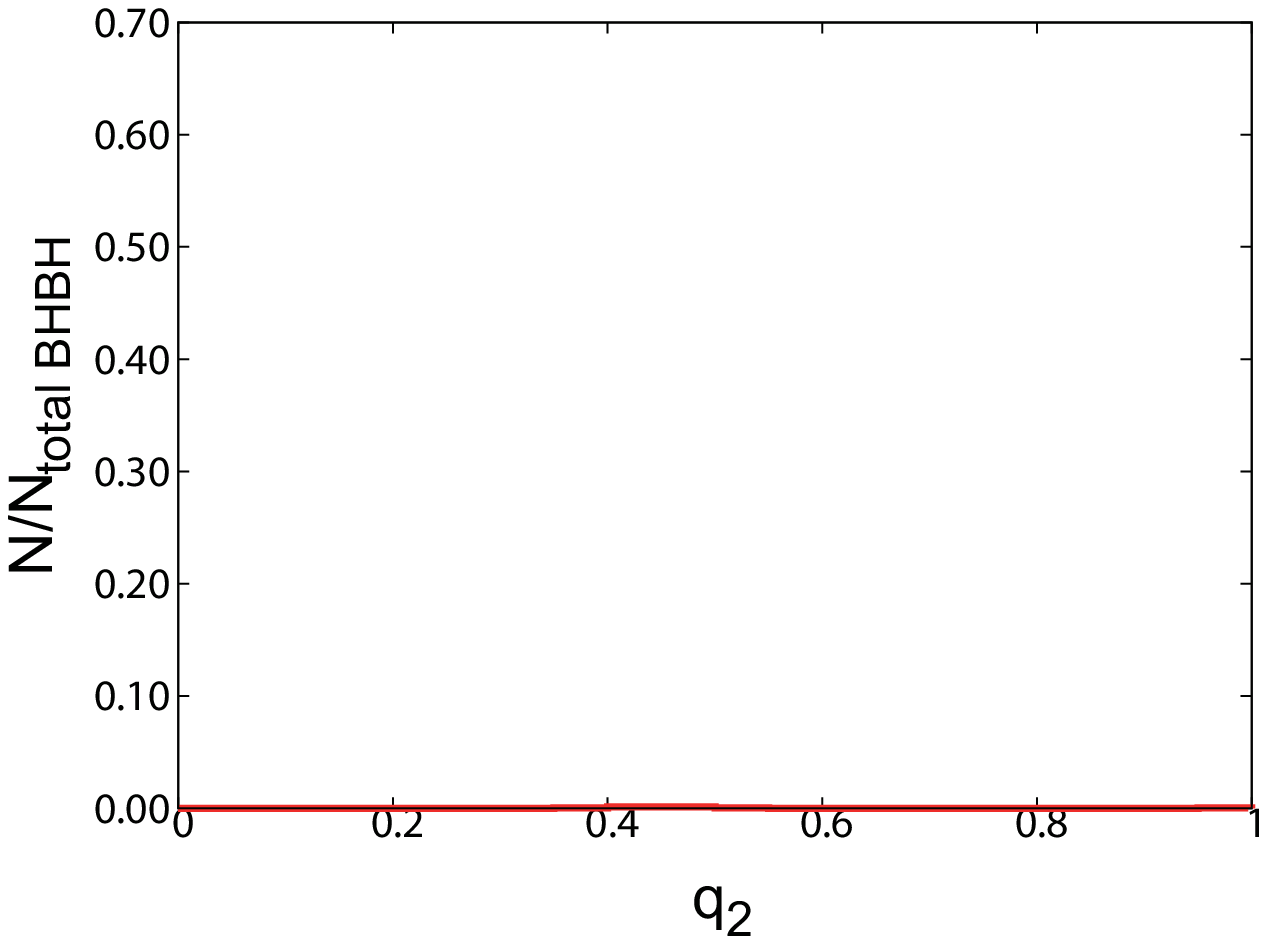}
        }
        \subfloat[$0.95<q_1<0.998$]{
            \includegraphics[scale=0.4]{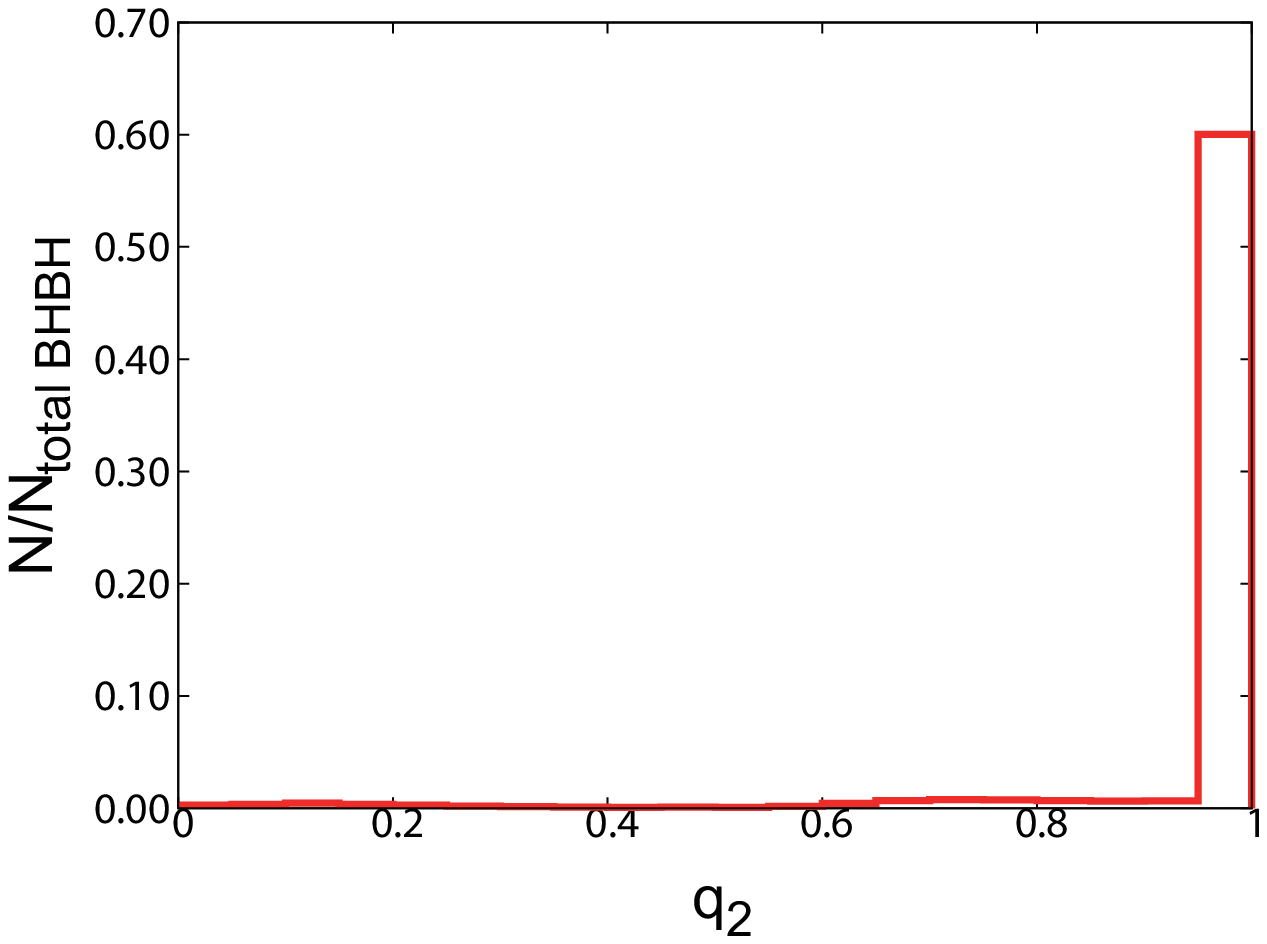}
        }
\end{center}
 \caption{(Top) The distribution of spin parameters for $\alpha\lambda=0.01$ model.
 $N_{\rm total \,BHBH}=27790$.
 (Bottom) Cross section views of distribution of spin parameter for $\alpha\lambda=0.01$ model.
 This figure is same as Figure~\ref{fig:spin1spin2standard} but for $\alpha\lambda=0.01$ model.}
 \label{fig:spin1spin2al=001}
\end{figure}

\begin{figure}[!ht]
\begin{center}
\includegraphics[width=0.5\textwidth,clip=true]{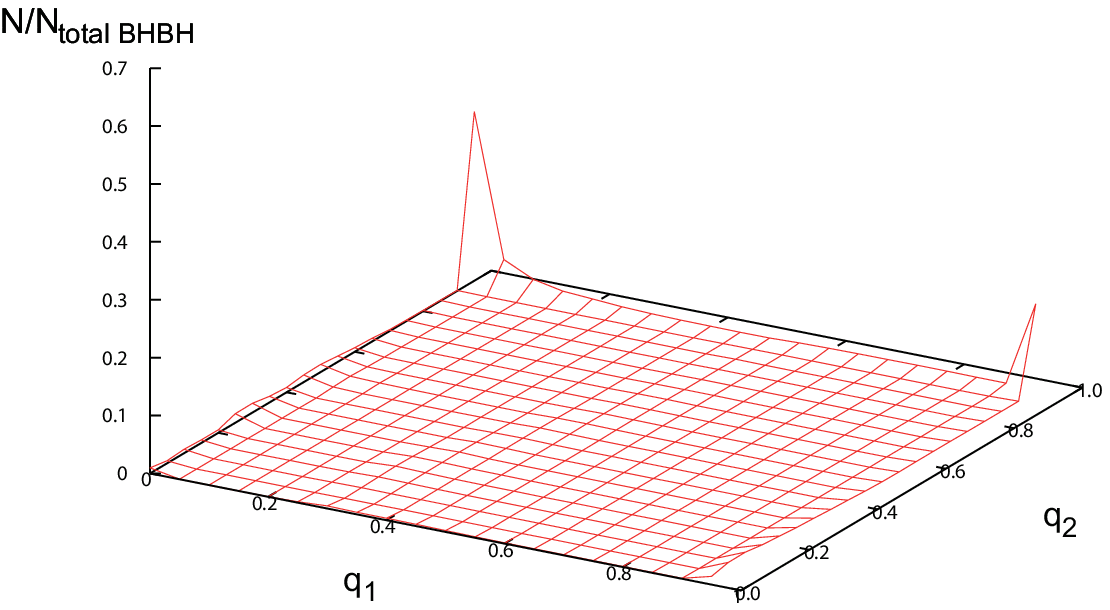}
        \subfloat[$0<q_1<0.05$]{
            \includegraphics[scale=0.4]{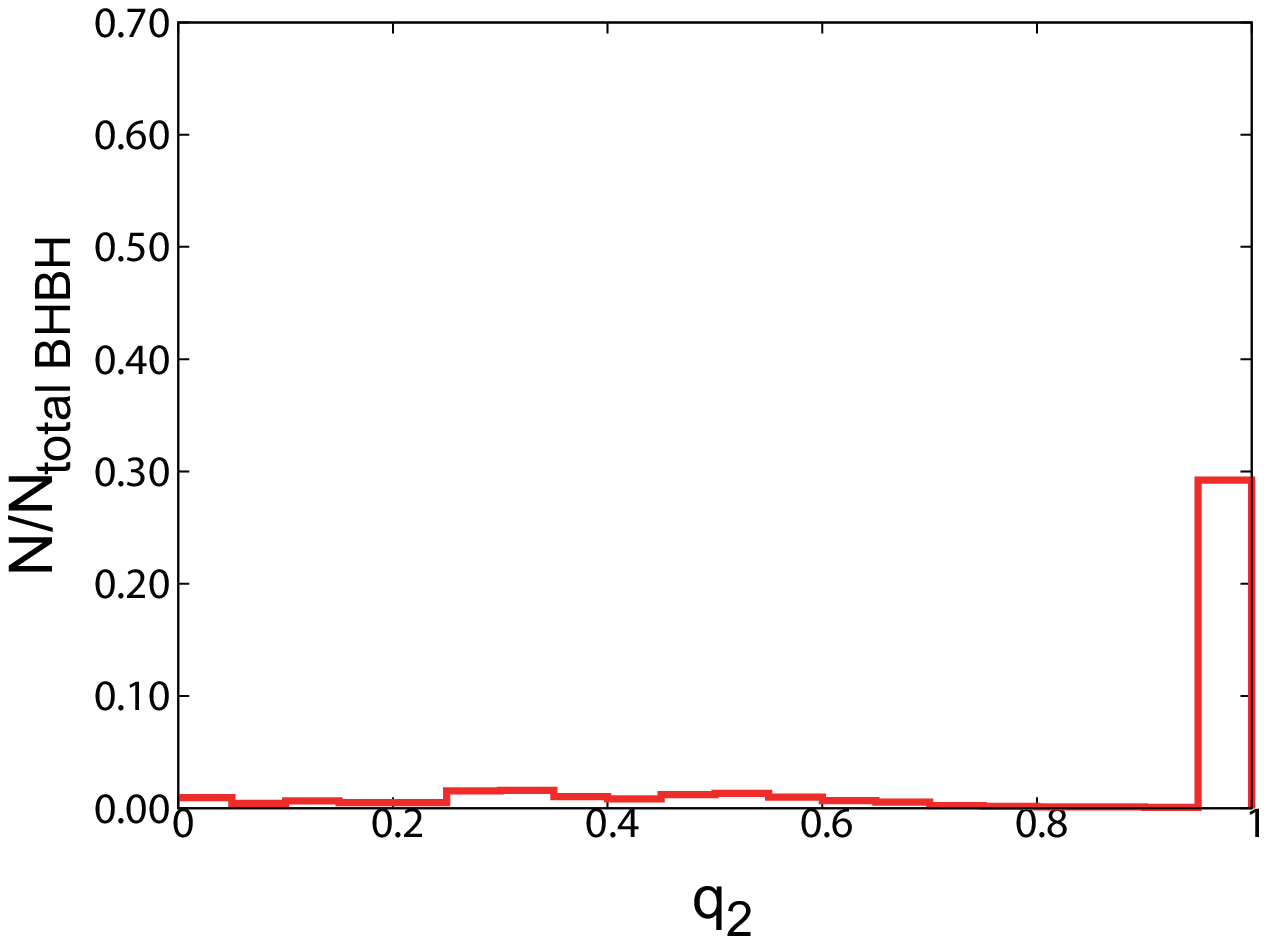}
        }
        \subfloat[$0.95<q_1<0.998$]{
            \includegraphics[scale=0.4]{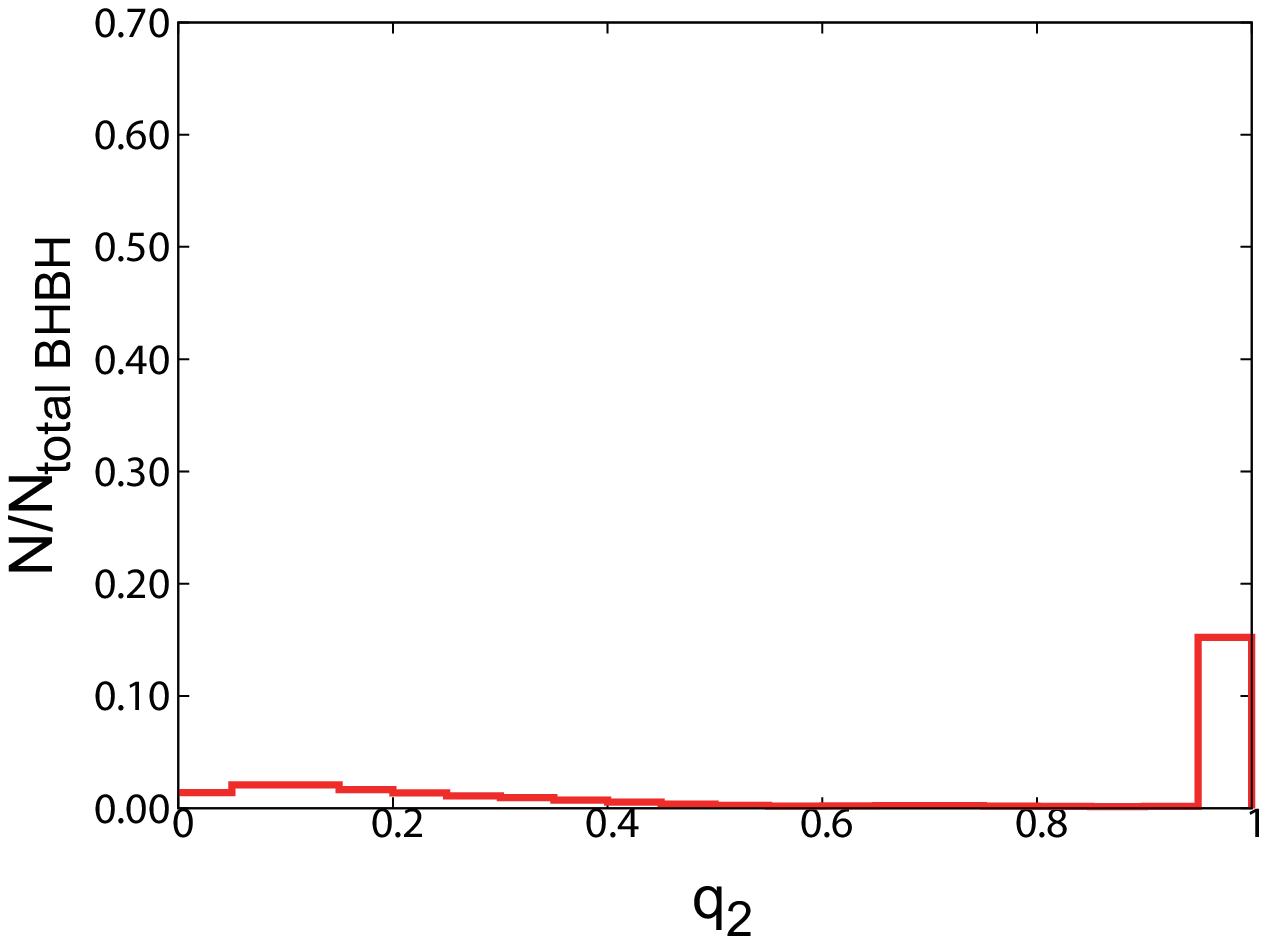}
        }
\end{center}
 \caption{(Top) The distribution of spin parameters for $\alpha\lambda=0.1$ model.
 $N_{\rm total \,BHBH}=124834$.
 (Bottom) Cross section views of distribution of spin parameter for $\alpha\lambda=0.1$ model.
 This figure is same as Figure~\ref{fig:spin1spin2standard} but for $\alpha\lambda=0.1$ model.}
 \label{fig:spin1spin2al=01}
\end{figure}

\begin{figure}[!ht]
\begin{center}
\includegraphics[width=0.5\textwidth,clip=true]{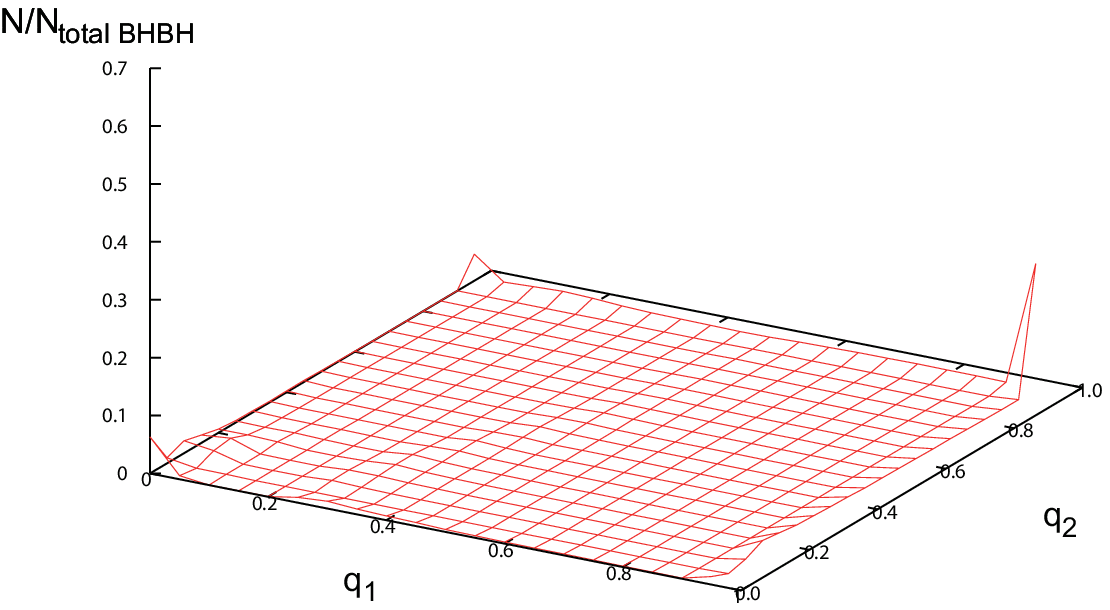}
        \subfloat[$0<q_1<0.05$]{
            \includegraphics[scale=0.4]{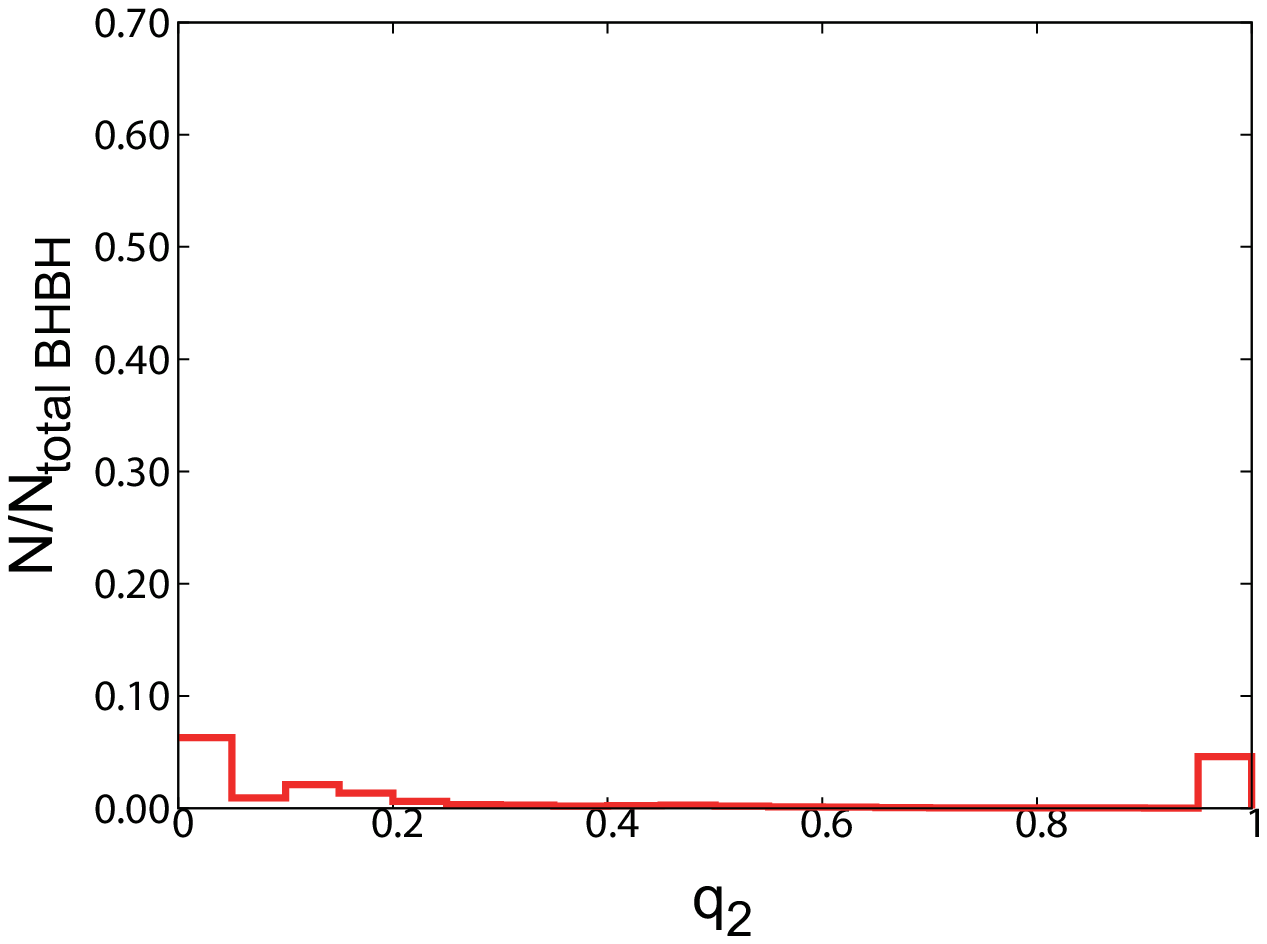}
        }
        \subfloat[$0.95<q_1<0.998$]{
            \includegraphics[scale=0.4]{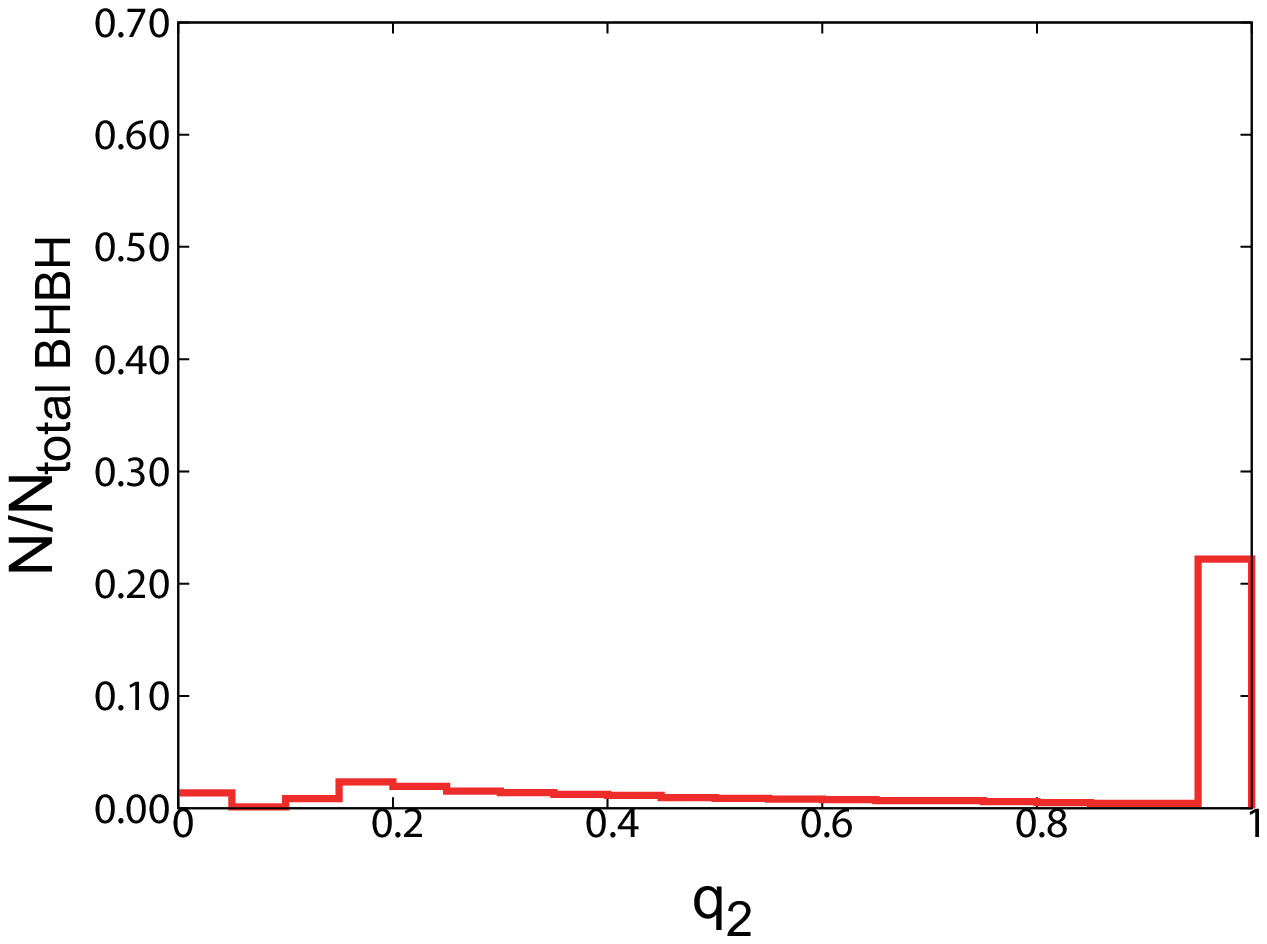}
        }
\end{center}
 \caption{(Top) The distribution of spin parameters for $\alpha\lambda=10$ model.
 $N_{\rm total \,BHBH}=93731$.
 (Bottom) Cross section views of distribution of spin parameter for $\alpha\lambda=10$ model.
 This figure is same as Figure~\ref{fig:spin1spin2standard} but for $\alpha\lambda=10$ model.}
 \label{fig:spin1spin2al=10}
\end{figure}

\begin{figure}[!ht]
\begin{center}
\includegraphics[width=0.5\textwidth,clip=true]{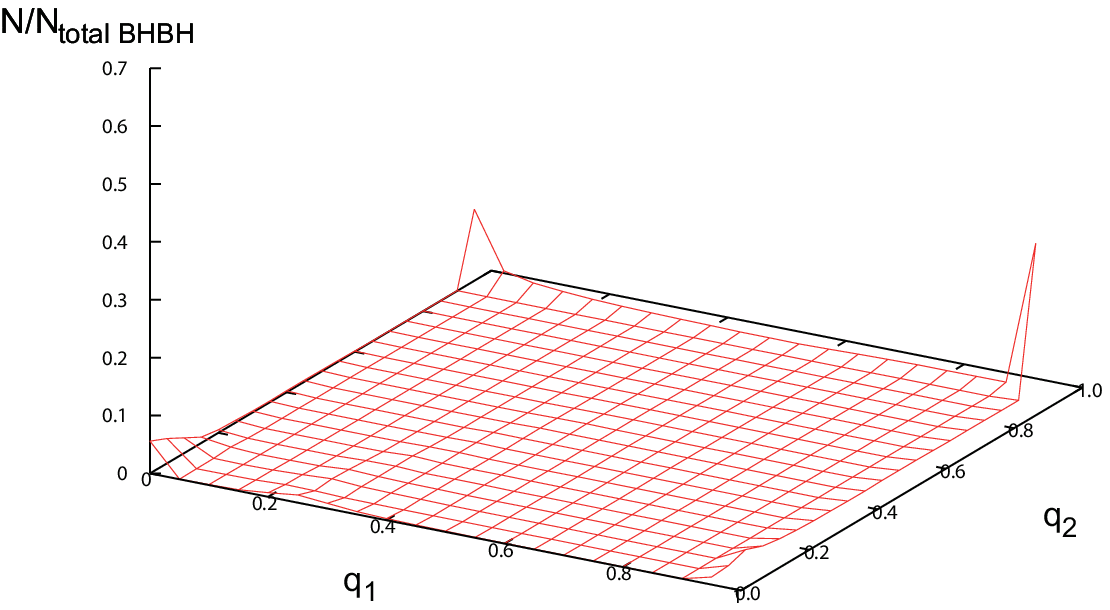}
        \subfloat[$0<q_1<0.05$]{
            \includegraphics[scale=0.4]{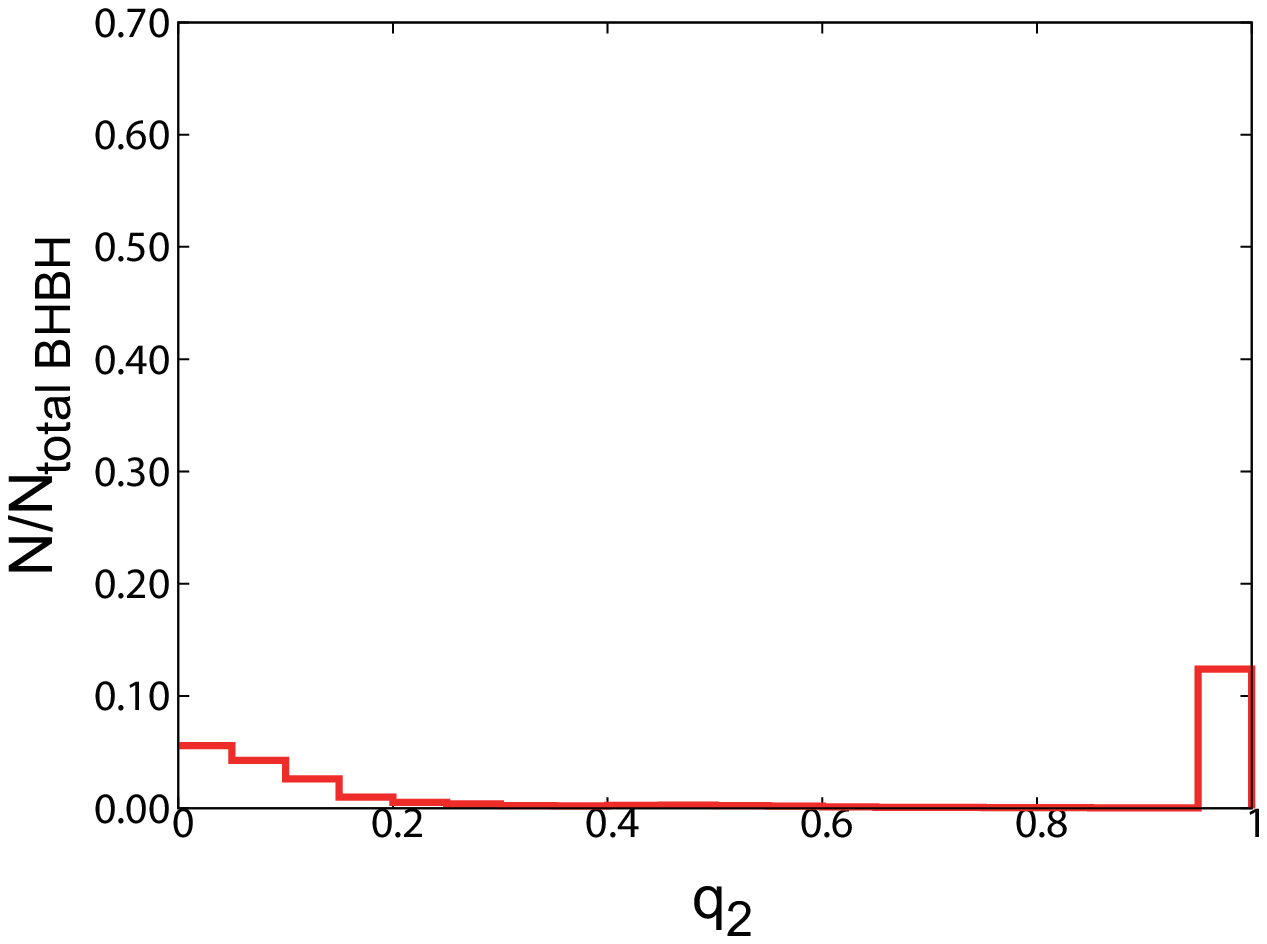}
        }
        \subfloat[$0.95<q_1<0.998$]{
            \includegraphics[scale=0.4]{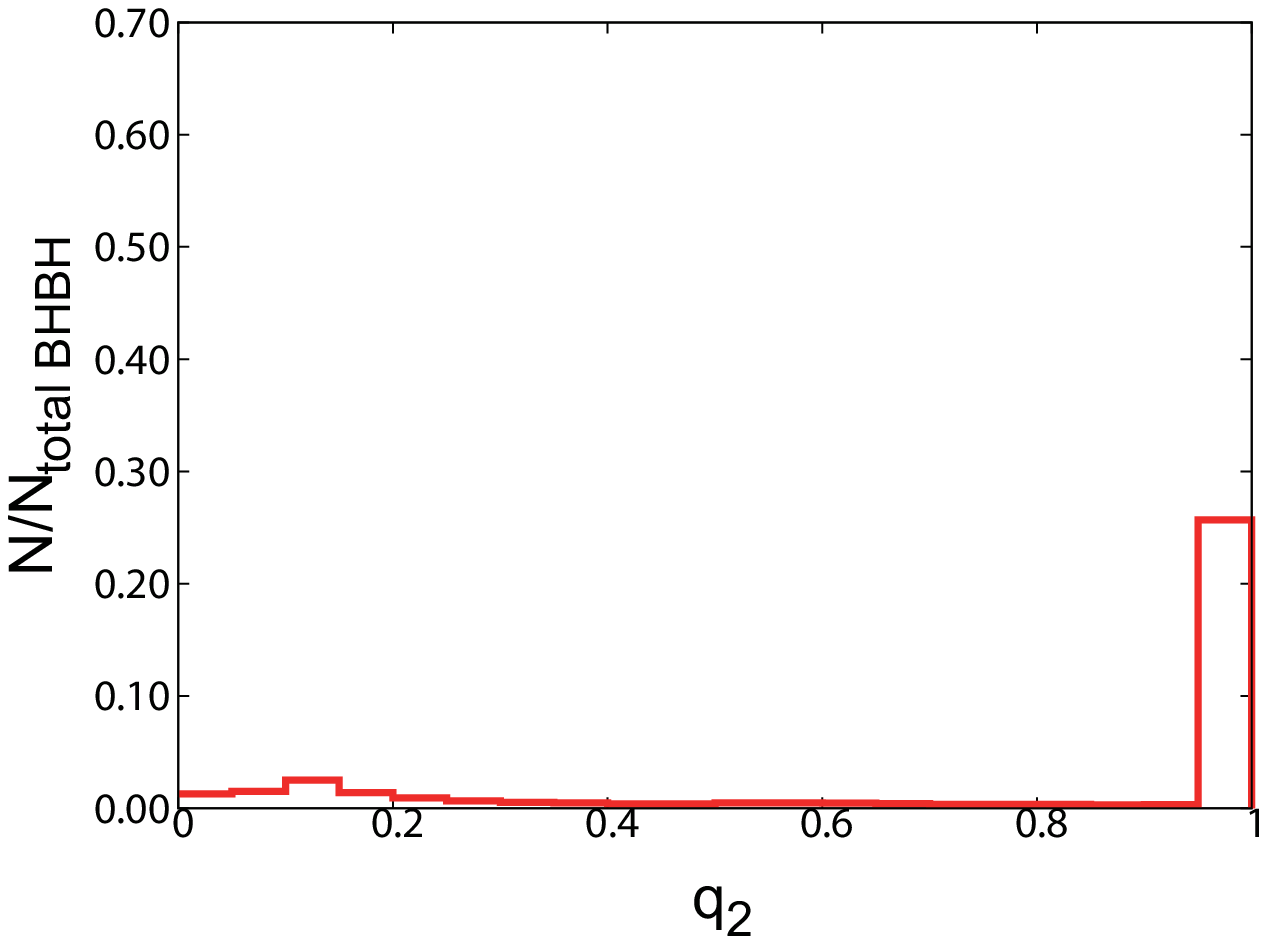}
        }
\end{center}
 \caption{(Top) The distribution of spin parameters for $\beta=0.5$ model.
 $N_{\rm total \,BHBH}=126093$.
 (Bottom) Cross section views of distribution of spin parameter for $\beta=0.5$ model.
 This figure is same as Figure~\ref{fig:spin1spin2standard} but for $\beta=0.5$ model.}
 \label{fig:spin1spin2b=05}
\end{figure}

\begin{figure}[!ht]
\begin{center}
\includegraphics[width=0.5\textwidth,clip=true]{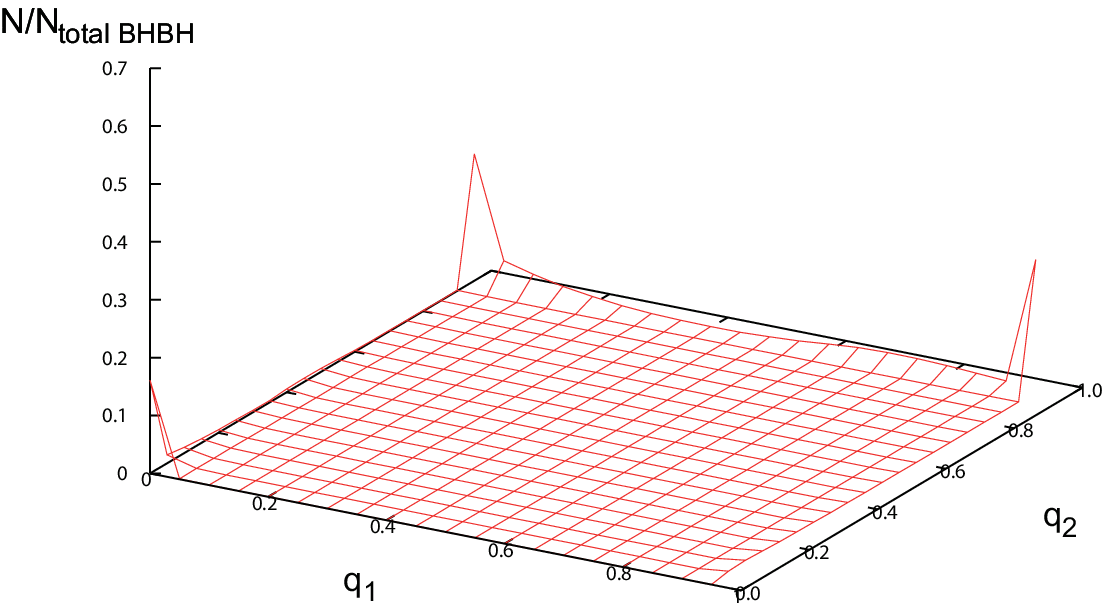}
        \subfloat[$0<q_1<0.05$]{
            \includegraphics[scale=0.4]{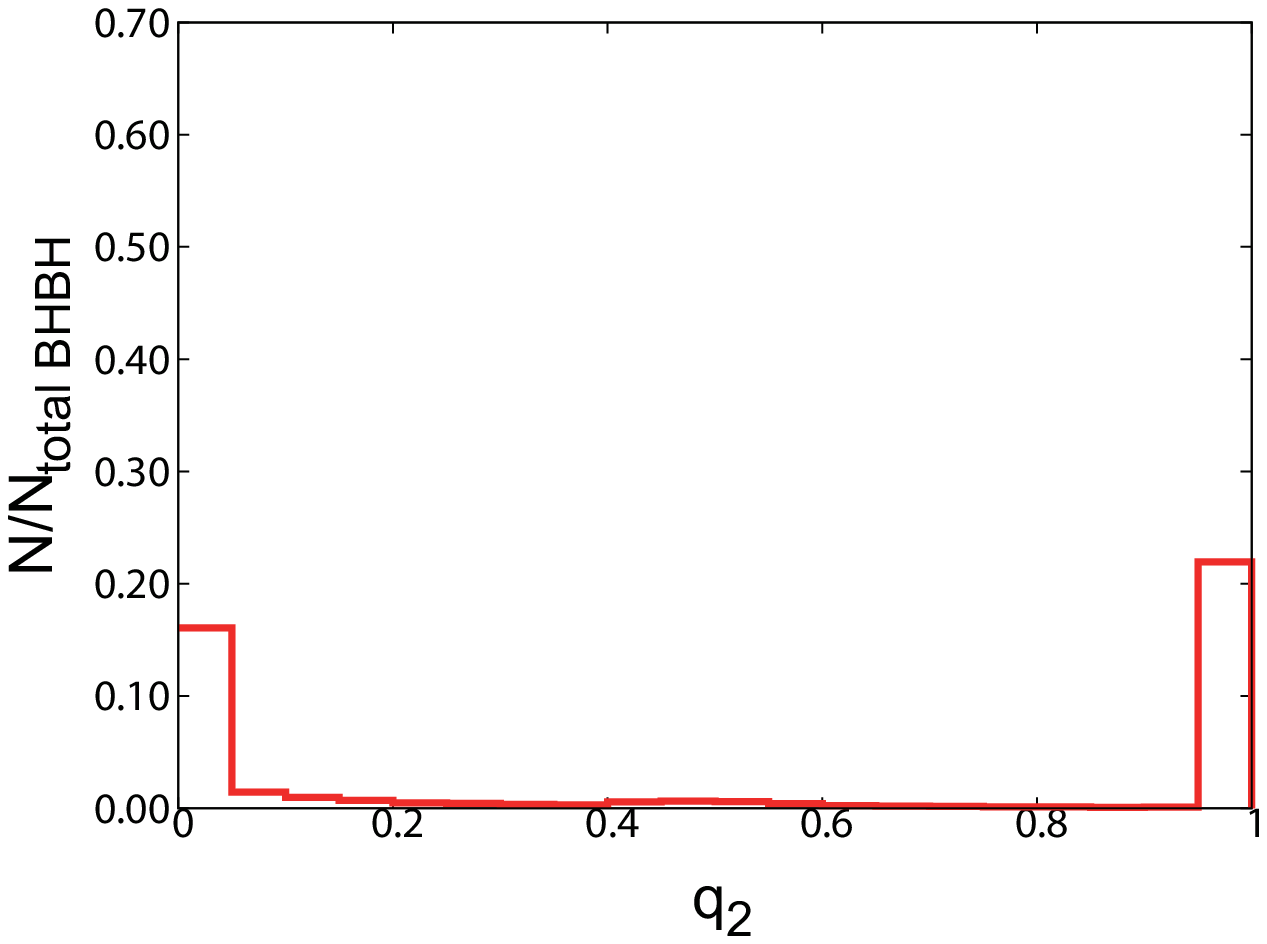}
        }
        \subfloat[$0.95<q_1<0.998$]{
            \includegraphics[scale=0.4]{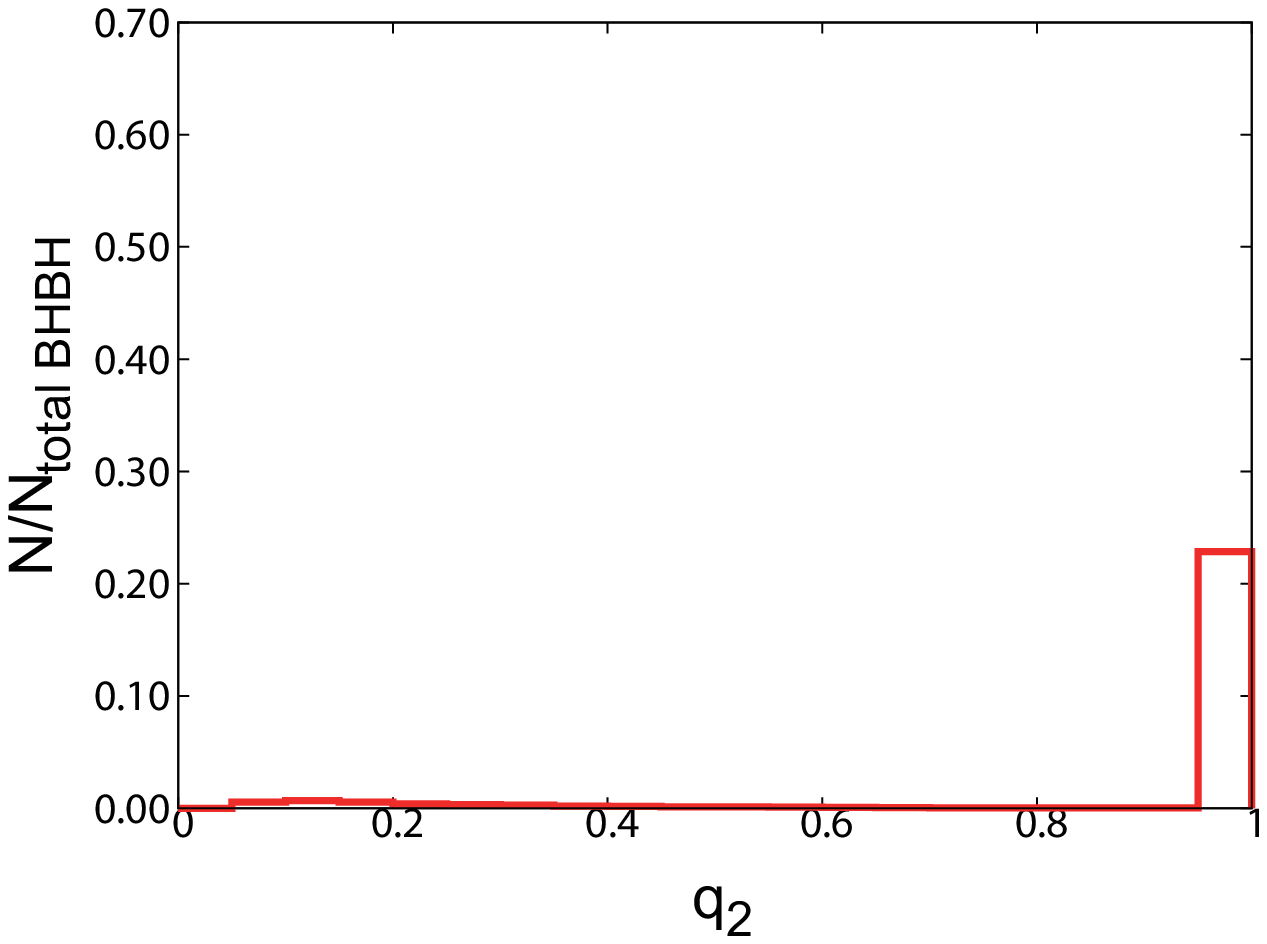}
        }
\end{center}
 \caption{(Top) The distribution of spin parameters for $\beta=1$ model.
 $N_{\rm total \,BHBH}=57028$.
 (Bottom) Cross section views of distribution of spin parameter for $\beta=1$ model.
 This figure is same as Figure~\ref{fig:spin1spin2standard} but for $\beta=1$ model.}
 \label{fig:spin1spin2b=1}
\end{figure}

Figures~\ref{fig:spin1spin2standard}--\ref{fig:spin1spin2b=1}
show the spin distributions of merging BBHs
and the cross section views of those spin distributions.
The spins of merging Pop III BBH can be roughly classified into three types:
group 1 in which both BHs have high spins $q\sim 0.998$,
group 2 in which both BHs have low spins, and
group 3 in which the one of the pair has the high spin $q\sim0.998$ and the other has low spin.

When the BH progenitor evolved via the CE phase, the Pop III BH has low spin,
and vice versa.
If the Pop III star which is a giant evolves via the CE phase,
the Pop III star loses the envelope and almost all the angular momentum
due to the envelope evaporation.
On the other hand, if the Pop III star evolves without the CE phase,
the Pop III star can have a high angular momentum.
Therefore, in the group 1 progenitors evolve without the CE phase
and the envelopes of the progenitors remain.
In the group 2, both stars evolve via the CE phases and
they lose their envelopes and almost all the angular momentum.
In the group 3, the primary evolves via the CE phase,
and the secondary evolves without the CE phase, or vice versa.

The IMF dependence of merging Pop III BBH spins is described as following
(see Figs.~\ref{fig:spin1spin2standard}, \ref{fig:spin1spin2logflat}
and \ref{fig:spin1spin2salpeter}).
The Pop III stars with masses $< 50 \msun$ evolve as a blue giant.
Thus, in the case of the IMF that light stars are majority such as Salpeter IMF,
the binaries tend to evolve only via the RLOF, not via the CE phase. 
Therefore, for the steeper IMF, we have larger numbers of merging Pop III BBHs
which have high spins.
Especially, in the case of Salpeter IMF
the BBHs whose spins are $q_1>0.95$ and $q_2>0.95$, are about $40\%$.

%%%%%%%%%%%%%%%%%%%%%%%%%%%%%%%%%%%%%%%%%%%%%%%%%%
%%%%%%%%%%%%%%%%%%%%%%%%%%%%%%%%%%%%%%%%%%%%%%%%%%IMF

%%%%%%%%%%%%%%%%%%%%%%%%%%%%%%%%%%%%%%%%%%%%%%%%%%

As for the IEF dependence, there is no tendency like the mass ratio distribution
(see Figs.~\ref{fig:spin1spin2standard}, 
\ref{fig:spin1spin2e=const} and \ref{fig:spin1spin2e-05}).

%%%%%%%%%%%%%%%%%%%%%%%%%%%%%%%%%%%%%%%%%%%%%%%%%%IEF

%%%%%%%%%%%%%%%%%%%%%%%%%%%%%%%%%%%%%%%%%%%%%%%%%%

The dependence on the CE parameter can be considered as the following
(see Figs.~\ref{fig:spin1spin2standard}, 
\ref{fig:spin1spin2al=001}, \ref{fig:spin1spin2al=01} and \ref{fig:spin1spin2al=10}).
In the $\alpha\lambda=0.01$ model, almost all merging Pop III BBHs have high spins. 
About $60\%$ of merging Pop III BBHs have $q_1>0.95$ and $q_2>0.95$ (i.e., group 1). 
This reason is that the progenitors which evolve via the CE phase
always merge during the CE phase due to too small $\alpha\lambda$.
Thus, the progenitors of merging Pop III BBHs in this model
evolve only via the RLOF and they do not lose the angular momentum via the CE phase. 
In the case of $\alpha\lambda=0.1$ model,
the fraction of group 2 is lower than that of our standard model
like the $\alpha\lambda=0.01$ model.
However, the fraction of group 1 is almost same as that of our standard model, 
and the fraction of group 3 is larger than that of our standard model
not like a $\alpha\lambda=0.01$ model.
This reason is that although the progenitors which become CE phases more than one,
merge during the CE phases due to small $\alpha\lambda$,
the progenitors which become the CE phase at once, 
do not merge during the CE phase,
and the Pop III BBHs which cannot merge within the Hubble time in our standard model, 
become to be able to merge within the Hubble time due to small $\alpha\lambda$.
In the $\alpha\lambda=10$ model, the shape of the spin distribution
is almost same as that of our standard model.
The difference of this model from our standard model is small increase
of the fraction of group 2
because the progenitors which merge during the CE phase in our standard model,
become to be able to survive due to large $\alpha\lambda$.

%%%%%%%%%%%%%%%%%%%%%%%%%%%%%%%%%%%%%%%%%%%%%%%%%%CE parameter

%%%%%%%%%%%%%%%%%%%%%%%%%%%%%%%%%%%%%%%%%%%%%%%%%%

As for the $\beta$ dependence
(see Figs.~\ref{fig:spin1spin2standard}, 
\ref{fig:spin1spin2b=05} and \ref{fig:spin1spin2b=1}),
Not only the stellar mass loss during the RLOF
but also the criterion of the dynamically unstable mass transfer such as a CE phase
are changed by $\beta$.
In $\beta=0.5$ model, the fraction of group 1 is larger than that of our standard model
because in this model the progenitors more hardly become the CE phase
than those of our standard model.
In the $\beta=1$ model,
the fraction of group 1 is larger than that of our standard model
like a $\beta=0.5$ model.
However, the fraction of group 1 is smaller than that of $\beta=0.5$ model
because in this model the progenitors lose a lot of angular momentum
during the RLOF due to high $\beta$.
In this model, since the mass transfer cannot become dynamically unstable, 
the progenitors of group 2 become the CE phase when the primary and secondary
become a giant at the same time, and merge.
On the other hand, the progenitors of group 3 become the CE phase
when the secondary merges the primary giant due to initial eccentricity.

%%%%%%%%%%%%%%%%%%%%%%%%%%%%%%%%%%%%%%%%%%%%%%%%%%beta
 
\clearpage

%%%%%%%%%%%%%%%%%%%%%%%%%%%%%%%%%%%%%%%%
\section{Remnant and event rate for ringdown gravitational waves}\label{sec:R_ER}
%%%%%%%%%%%%%%%%%%%%%%%%%%%%%%%%%%%%%%%%

%%%%%%%%%%
\subsection{Remnant mass and spin}

\begin{figure}[!ht]
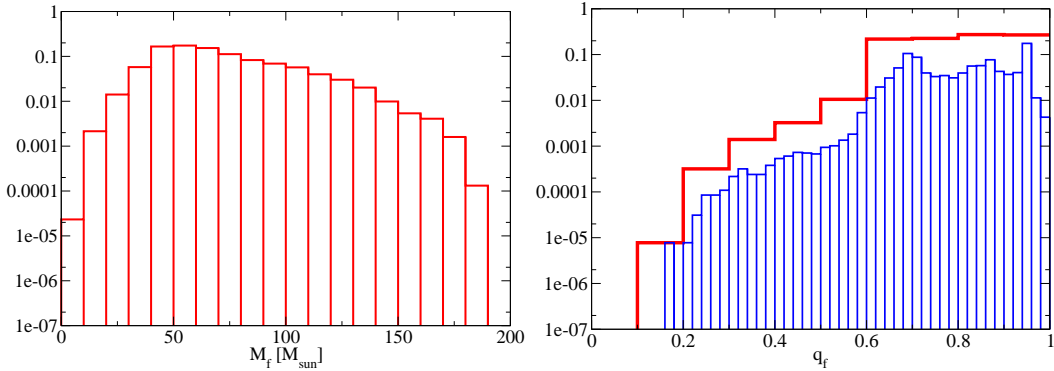

\begin{center}
 \includegraphics[width=0.45\textwidth,clip=true]{./figs/Mrem_standard}
 \includegraphics[width=0.45\textwidth,clip=true]{./figs/qrem_standard}
\end{center}
 \caption{The remnant mass $M_f$ (left)
 and spin $q_f$ (right) for our standard model.}
 \label{fig:rem_standard}
\end{figure}

\begin{figure}[!ht]
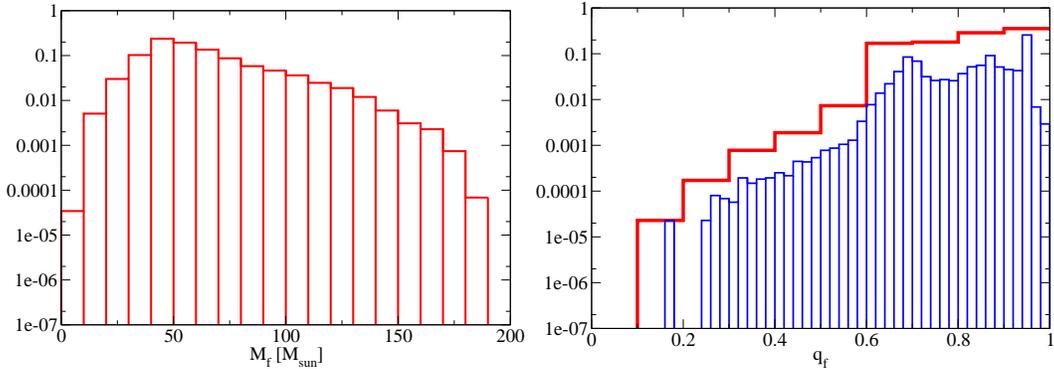

\begin{center}
 \includegraphics[width=0.45\textwidth,clip=true]{./figs/Mrem_M-1}
 \includegraphics[width=0.45\textwidth,clip=true]{./figs/qrem_M-1}
\end{center}
 \caption{The remnant mass $M_f$ (left)
 and spin $q_f$ (right) for IMF: logflat model.}
 \label{fig:rem_logflat}
\end{figure}

\begin{figure}[!ht]
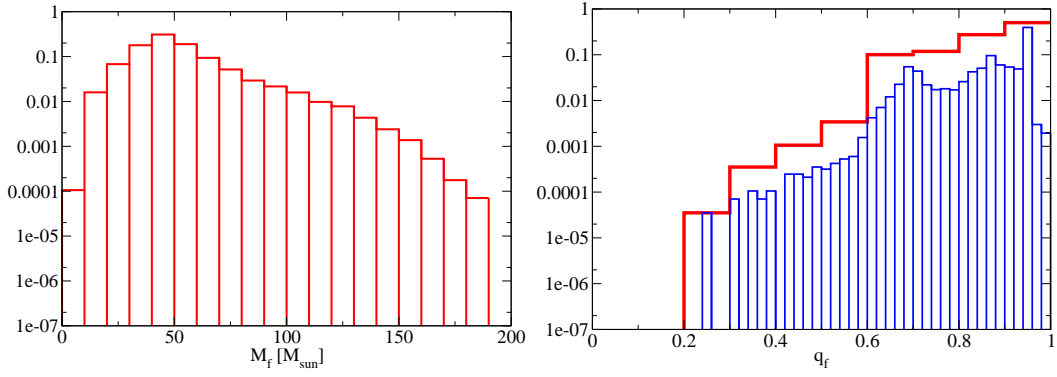

\begin{center}
 \includegraphics[width=0.45\textwidth,clip=true]{./figs/Mrem_Salpeter}
 \includegraphics[width=0.45\textwidth,clip=true]{./figs/qrem_Salpeter}
\end{center}
 \caption{The remnant mass $M_f$ (left)
 and spin $q_f$ (right) for IMF: Salpeter model.}
 \label{fig:rem_salpeter}
\end{figure}

\begin{figure}[!ht]
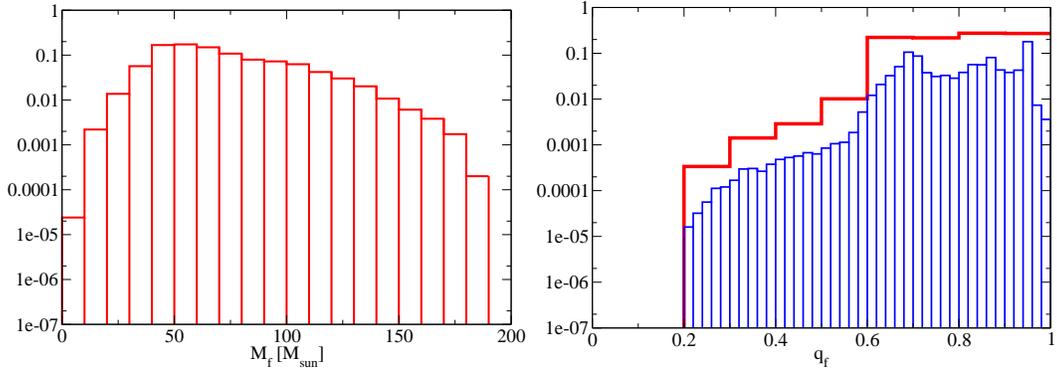

\begin{center}
 \includegraphics[width=0.45\textwidth,clip=true]{./figs/Mrem_e=const.eps}
 \includegraphics[width=0.45\textwidth,clip=true]{./figs/qrem_e=const.eps}
\end{center}
 \caption{The remnant mass $M_f$ (left)
 and spin $q_f$ (right) for IEF: $e=$const model.}
 \label{fig:rem_e=const}
\end{figure}

\begin{figure}[!ht]
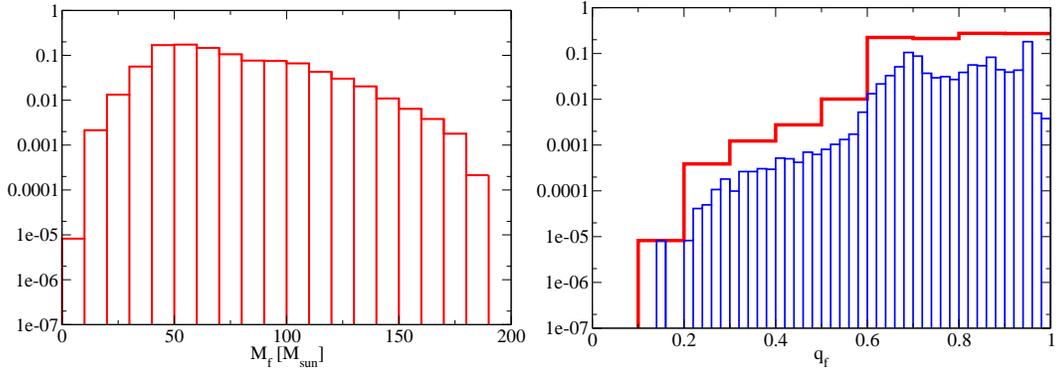

\begin{center}
 \includegraphics[width=0.45\textwidth,clip=true]{./figs/Mrem_e-0.5.eps}
 \includegraphics[width=0.45\textwidth,clip=true]{./figs/qrem_e-0.5.eps}
\end{center}
 \caption{The remnant mass $M_f$ (left)
 and spin $q_f$ (right) for IEF: $e^{-0.5}$ model.}
 \label{fig:rem_e-0.5}
\end{figure}

\begin{figure}[!ht]
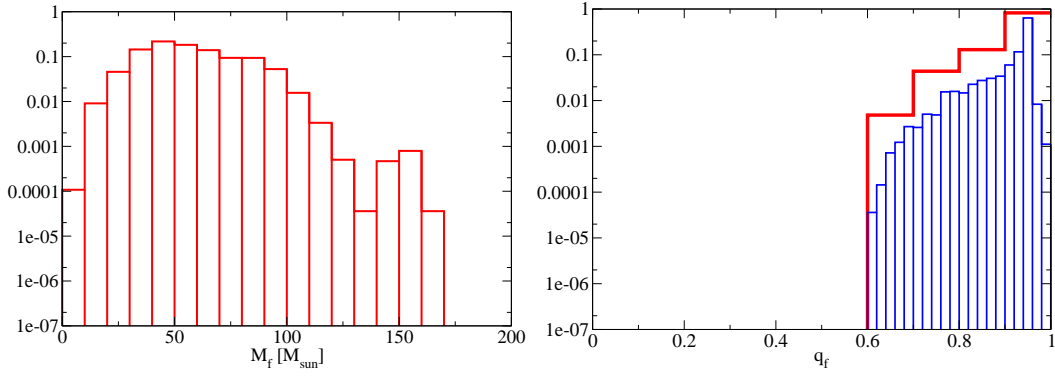

\begin{center}
 \includegraphics[width=0.45\textwidth,clip=true]{./figs/Mrem_al=0.01.eps}
 \includegraphics[width=0.45\textwidth,clip=true]{./figs/qrem_al=0.01.eps}
\end{center}
 \caption{The remnant mass $M_f$ (left)
 and spin $q_f$ (right) for $\alpha \lambda=0.01$ model.}
 \label{fig:rem_al001}
\end{figure}

\begin{figure}[!ht]
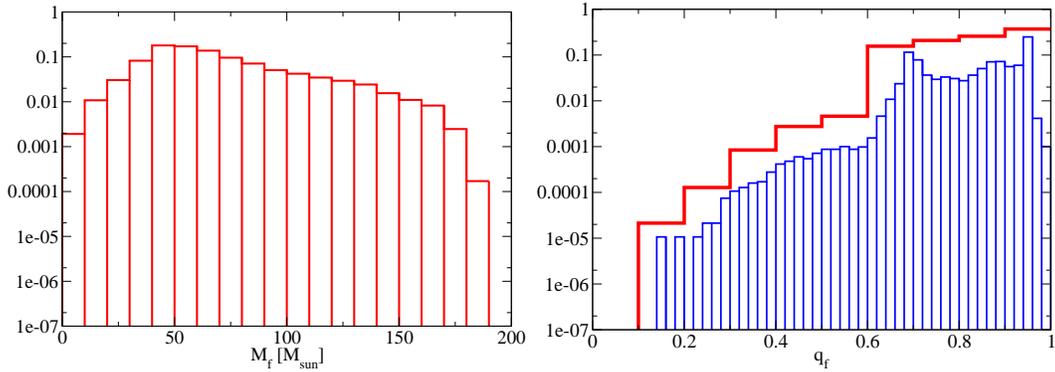

\begin{center}
 \includegraphics[width=0.45\textwidth,clip=true]{./figs/Mrem_al=10.eps}
 \includegraphics[width=0.45\textwidth,clip=true]{./figs/qrem_al=10.eps}
\end{center}
 \caption{The remnant mass $M_f$ (left)
 and spin $q_f$ (right) for $\alpha \lambda=10$ model.}
 \label{fig:rem_al10}
\end{figure}

\begin{figure}[!ht]
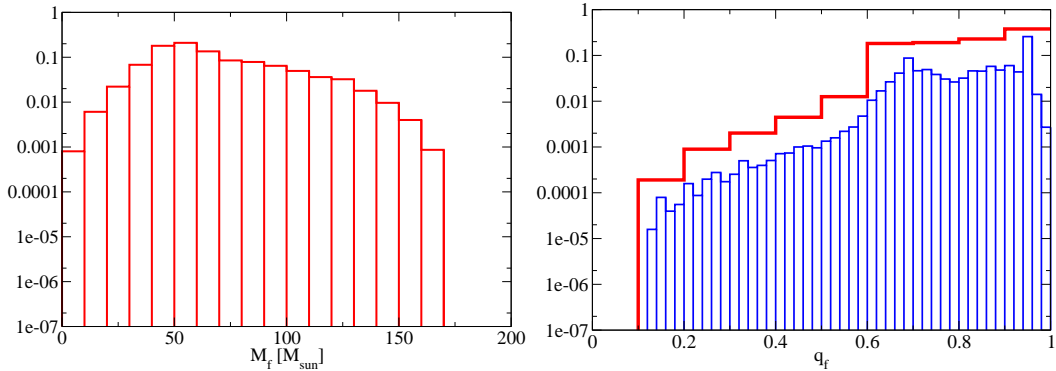

\begin{center}
 \includegraphics[width=0.45\textwidth,clip=true]{./figs/Mrem_b=0.5.eps}
 \includegraphics[width=0.45\textwidth,clip=true]{./figs/qrem_b=0.5.eps}
\end{center}
 \caption{The remnant mass $M_f$ (left)
 and spin $q_f$ (right) for $\beta=0.5$ model.}
 \label{fig:rem_b=05}
\end{figure}

\begin{figure}[!ht]
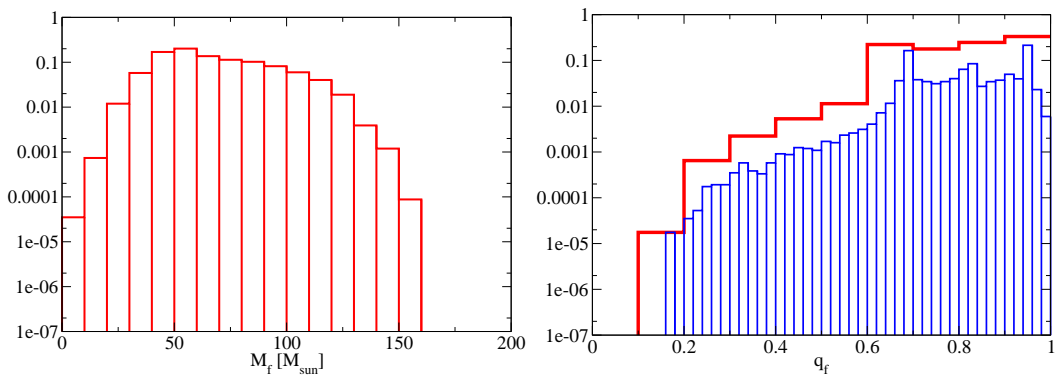

\begin{center}
 \includegraphics[width=0.45\textwidth,clip=true]{./figs/Mrem_b=1.eps}
 \includegraphics[width=0.45\textwidth,clip=true]{./figs/qrem_b=1.eps}
\end{center}
 \caption{The remnant mass $M_f$ (left)
 and spin $q_f$ (right) for $\beta=1$ model.}
 \label{fig:rem_b=1}
\end{figure}

Based on Ref.~\cite{Healy:2014yta}
(see also Refs.~\cite{Barausse:2012qz,Hofmann:2016yih}),
we calculate the remnant mass $M_f$
and spin $q_f$ from given BH binary parameters,
$M_1$, $M_2$, $q_1$ and $q_2$
(see Ref.~\cite{Kinugawa:2016mfs} for the detailed discussion).
The remnant mass and spin for each case is show
in Figs.~\ref{fig:rem_standard}--\ref{fig:rem_b=1}.
Here, we have normalized the distribution, and used
binning with $\Delta M_f=10\,\msun$ for $M_f$
and $\Delta q_f=0.1$ (thick, red) and $0.02$ (thin, blue) for $q_f$.

The IMF dependence shown in Figs.~\ref{fig:rem_standard},
\ref{fig:rem_logflat} and \ref{fig:rem_salpeter}
is described as following for the remnant mass and spin.
When we treat the steeper IMF, we have a lower number
of high mass remnants. On the other hand, the number of high spin remnants
increase slightly in the steeper IMF cases.
This is because in the steeper IMF models
we have a large number of progenitors with mass smaller than $50~\msun$.

%%%%%%%%%%%%%%%%%%%%%%%%%%%%%%%%%%%%%%%%
%%%%%%%%%%%%%%%%%%%%%%%%%%%%%%%%%%%%%%%%

As for the IEF dependence, we find that 
in Figs.~\ref{fig:rem_standard}, \ref{fig:rem_e=const} and \ref{fig:rem_e-0.5}
there is no strong tendency.

%%%%%%%%%%%%%%%%%%%%%%%%%%%%%%%%%%%%%%%%
%%%%%%%%%%%%%%%%%%%%%%%%%%%%%%%%%%%%%%%%

Next, from Figs.~\ref{fig:rem_standard}, \ref{fig:rem_al001}, \ref{fig:rem_al01} and \ref{fig:rem_al10}
the CE parameter dependence is described as following.
In the $\alpha\lambda=0.01$ model,
the maximum of the remnant mass becomes much smaller
than that of our standard model.
This is because the high mass progenitors merge
during a CE phase due to too small $\alpha\lambda$,
As for the remnant spin, 
we do not have remnant spins which are smaller than $0.55$
since BBHs tend to be equal mass.
If a light BH falls into a nonspining massive BH,
the remnant BH can have a small spin ($q_f < 0.6$).
However, in the above model many BBHs are equal mass.
In the $\alpha\lambda=0.1$ model, the maximum remnant mass is
smaller than that of our standard model again.
In this model, the fraction of the remnant spins with $0.7<q_f<0.8$ is
larger than that for our standard model because the fraction of group 3
in this model is larger than that of our standard model.
In the $\alpha\lambda=10$ model, the maximum remnant mass
is larger than that of our standard model.
This is because the high mass progenitors which tend to become a CE phase,
can survive more easily during the CE phase
than those of our standard model.

%%%%%%%%%%%%%%%%%%%%%%%%%%%%%%%%%%%%%%%%

\begin{figure}[!ht]
\begin{center}
 \includegraphics[width=0.45\textwidth,clip=true]{./figs/Mrem_al=0.1.eps}
 \includegraphics[width=0.45\textwidth,clip=true]{./figs/qrem_al=0.1.eps}
\end{center}
 \caption{The remnant mass $M_f$ (left)
 and spin $q_f$ (right) for $\alpha \lambda=0.1$ model.}
 \label{fig:rem_al01}
\end{figure}

%%%%%%%%%%%%%%%%%%%%%%%%%%%%%%%%%%%%%%%%

As for the $\beta$ dependence
we find that from Figs.~\ref{fig:rem_standard}, \ref{fig:rem_b=05} and \ref{fig:rem_b=1}
the maximum of remnant mass becomes lower for the higher $\beta$,
due to the mass loss during a RLOF.

%%%%%%%%%%%%%%%%%%%%%%%%%%%%%%%%%%%%%%%%
%%%%%%%%%%%%%%%%%%%%%%%%%%%%%%%%%%%%%%%%

%%%%%%%%%%
\subsection{Event rates for ringdown gravitational waves}

To estimate the event rate for ringdown gravitational waves, it is necessary to have
the merger rate density of Pop III BBHs. 
The merger rate density $R_m~[{\rm Myr^{-1} Mpc^{-3}}]$
has been derived for various models
in Ref.~\cite{Kinugawa:2015nla},
and can be approximated by a fitting formula for low redshift.
This is summarized in Table~\ref{tab:MRDfit}.
In practice, we have considered the fitting for $R_m$
in terms of the redshift $z$ up to $z =2$, but
the above $R_m$ is derived by using $z \propto D$
where $D$ denotes the (luminosity) distance,
because we use it only up to $z \sim 0.2$ in this paper.

\begin{table}[!t]
\caption{Fitting formulas of the merger rate density
 $R_m~[{\rm Myr^{-1} Mpc^{-3}}]$ for low redshift in the center column.
 Here, $D$ denotes the (luminosity) distance.
 In the right column, the event rates $[{\rm yr}^{-1}]$ divided by dependence
 on the star formation rate ${\rm SFR_p}$ and the fraction of the binary ${\rm f_b}$
 are shown for each model.
 We consider the events with ${\rm SNR}>8$ for the KAGRA detector here.
}
\label{tab:MRDfit}
\begin{center}
\begin{tabular}{|c||c|c|}
\hline
Model & $R_m~[{\rm Myr^{-1} Mpc^{-3}}]$ & ${\rm SNR}>8$ \\
\hline
Standard & $0.024+0.0080~(D/1\,{\rm Gpc})$ & 446 \\
IMF: logflat & $0.023+0.0064~(D/1\,{\rm Gpc})$ & 255 \\
IMF: Salpeter & $0.014+0.0034~(D/1\,{\rm Gpc})$ & 61.5 \\
IEF: $e=$const & $0.023+0.0075~(D/1\,{\rm Gpc})$ & 452 \\
IEF: $e^{-0.5}$ &$0.022+0.0071~(D/1\,{\rm Gpc})$ & 451 \\
$\alpha \lambda = 0.01$ & $0.0024+0.0018~(D/1\,{\rm Gpc})$ & 5.87 \\
$\alpha \lambda = 0.1$ & $0.019+0.0089~(D/1\,{\rm Gpc})$ & 146 \\
$\alpha \lambda = 10$ & $0.017+0.0066~(D/1\,{\rm Gpc})$ & 372 \\
$\beta = 0.5$ & $0.030+0.0099~(D/1\,{\rm Gpc})$ & 499 \\
$\beta = 1$ & $0.017+0.0029~(D/1\,{\rm Gpc})$ & 158 \\
\hline
\end{tabular}
\end{center}
\end{table}

Using Ref.~\cite{Flanagan:1997sx}, 
we calculate the angle averaged signal-to-noise ratio (SNR) as
\bea
{\rm SNR} =
\sqrt{\frac{8}{5}}
\frac{4 \eta}{F(q_f)} \sqrt{\frac{\epsilon_r M_f}{S_n(f_c)}}\frac{M_f}{D} \,,
\label{eq:SNR}
\eea
where we assume $\epsilon_r=3\%$ of the total mass energy radiated in the ringdown phase.
Note that for simplicity, any effect of the cosmological distance is ignored here.
The symmetric mass ratio $\eta = M_1M_2/(M_1+M_2)^2$
is evaluated from the inspiral phase, and 
\bea
F(q_f) = 1.5251 - 1.1568 (1-q_f)^{0.1292} \,,
\quad
f_c = \frac{1}{2\pi M_{f}} F(q_f) \,,
\eea
are obtained from the remnant BH's mass and spin
(see Ref.~\cite{Berti:2005ys}).
We evaluate the above SNR of the QNM GWs
in the expected KAGRA noise curve $S_n(f)$~\cite{Somiya:2011np,Aso:2013eba} 
[bKAGRA, VRSE(D) configuration]
(see Ref.~\cite{Kinugawa:2016mfs} for the detailed calculation).
This noise curve is presented in Ref.~\cite{bKAGRA},
and we use the fitting noise curve obtained in Ref.~\cite{Nakano:2015uja},
based on Ref.~\cite{bKAGRA}.
Then, the event rate for a given SNR
is derived by using the merger rate density in Table~\ref{tab:MRDfit}.
In the right column of Table~\ref{tab:MRDfit},
we present the event rate with SNR$>8$.
Here, the event rates $[{\rm yr}^{-1}]$ have been
divided by dependence on the star formation rate
${\rm SFR_p}$ and the fraction of the binary ${\rm f_b}$

Next, using the event with SNR$>35$,
we summarize various event rates in
Tables~\ref{tab:detection_all_q} and \ref{tab:detection_C}.
Table~\ref{tab:detection_all_q} shows
the total event rates $[{\rm yr}^{-1}]$
divided by dependence on the star formation rate
${\rm SFR_p}$ and the fraction of the binary ${\rm f_b}$
for $10$ models, and those for the remnant BH with $q_f > 0.7$ and $0.95$.

In Table~\ref{tab:detection_C},
we present the detection rate $[{\rm yr}^{-1}]$ divided
by dependence on the star formation rate ${\rm SFR_p}$
and the fraction of the binary ${\rm f_b}$
as a function of the lower limit of 
the solid angle of a sphere $4\pi C$
by which we can estimate the contribution of the ergoregion. 
The relation between this $C$ and the spin parameter $q$
was obtained in Ref.~\cite{Nakano:2016sgf}
(see also resent studies~\cite{Nakamura:2016gri,Nakano:2016sgf,
Nakamura:2016yjl,Nakano:2016zvv}).

\begin{table}[!t]
\caption{
 The total event rate $[{\rm yr}^{-1}]$ divided by dependence on the star formation rate
 ${\rm SFR_p}$ and the fraction of the binary ${\rm f_b}$
 for each model and those for the final $q_f > 0.7$ and $0.95$ BHs
 in the case of ${\rm SNR}>35$ for the KAGRA detector.
}
\label{tab:detection_all_q}
\begin{center}
\begin{tabular}{|c||c|c|c|}
\hline
Model &
all 
& $0.7<q_f$
& $0.95<q_f$ \\
\hline 
Standard &
3.73 & 2.45
& 0.0260 \\
\hline
IMF: logflat&
2.17 & 1.41
& 0.0152 \\
\hline
IMF: Salpeter &
0.530 & 0.337
& 0.00300 \\
\hline
IEF: $e=$const &
3.80 & 2.38
& 0.0248 \\
\hline
IEF: $e^{-0.5}$ &
3.82 & 2.38
& 0.0228 \\
\hline
$\alpha \lambda = 0.01$ &
0.0463 & 0.0411
& 0.00268 \\
\hline
$\alpha \lambda = 0.1$ &
1.23 & 0.864
& 0.0114 \\
\hline
$\alpha \lambda = 10$ &
2.96 & 2.16
& 0.0372 \\
\hline
$\beta = 0.5$ &
4.21 & 3.29
& 0.0157 \\
\hline
$\beta = 1$ &
1.58 & 0.705
& 0.0128 \\
\hline
\end{tabular}
\end{center}
\end{table}

\begin{table}[!t]
\caption{
 The event rates $[{\rm yr}^{-1}]$ divided by dependence on the star formation rate
 ${\rm SFR_p}$ and the fraction of the binary ${\rm f_b}$ as a function 
 of the lower limit of the solid angle of a sphere $4\pi C$
 where the QNM is mainly emitted from the ergoregion,
 in the case of SNR$>35$ for the KAGRA detector.
}
\label{tab:detection_C}
\begin{center}
\begin{tabular}{|c||c|c|c|c|c|c|}
\hline
Model &
$0.5<C$ & $0.7<C$ & $0.9<C$ & $0.95<C$ & $0.97<C$ & $0.99<C$ \\
\hline 
Standard &
2.23 & 1.10 & 0.356 & 0.162 & 0.117 & 0.0780 \\
\hline
IMF: logflat &
1.29 & 0.621 & 0.207 & 0.102 & 0.0683 & 0.0454 \\
\hline
IMF: Salpeter &
0.309 & 0.145 & 0.0489 & 0.0234 & 0.0156 & 0.00960 \\
\hline
IEF: $e=$const &
2.18 & 0.998 & 0.288 & 0.132 & 0.0955 & 0.0621 \\
\hline
IEF: $e^{-0.5}$ &
2.15 & 0.936 & 0.237 & 0.115 & 0.0825 & 0.0518 \\
\hline
$\alpha \lambda = 0.01$ &
0.0399 & 0.0318 & 0.0131 & 0.00966 & 0.00839 & 0.00503 \\
\hline
$\alpha \lambda = 0.1$ &
0.820 & 0.391 & 0.135 & 0.0609 & 0.0390 & 0.0222 \\
\hline
$\alpha \lambda = 10$ &
2.01 & 1.09 & 0.470 & 0.307 & 0.246 & 0.187 \\
\hline
$\beta = 0.5$ &
3.09 & 0.979 & 0.294 & 0.127 & 0.0990 & 0.0292 \\
\hline
$\beta = 1$ &
0.666 & 0.405 & 0.0278 & 0.0218 & 0.0187 & 0.0141 \\
\hline
\end{tabular}
\end{center}
\end{table}

%%%%%%%%%%%%%%%%%%%%%%%%%%%%%%%%%%%%%%%%
\section{Discussions}\label{sec:dis}
%%%%%%%%%%%%%%%%%%%%%%%%%%%%%%%%%%%%%%%%

In this paper, we extended our previous work~\cite{Kinugawa:2016mfs}
(standar model in this paper) with various parameter dependence
of the Pop III binary population synthesis calculation.
As shown in the right column of Table~\ref{tab:MRDfit},
the detection rate with SNR$>8$
for the second generation GW detectors such as KAGRA
was obtained as
$5.9-500~{\rm events~yr^{-1}}~({\rm SFR_p}/ (10^{-2.5}~M_\odot~{\rm yr^{-1}~Mpc^{-3}})) 
\cdot ({\rm [f_b/(1+f_b)]/0.33})$ between 10 models.
Recently, Kushnir et al.~\cite{Kushnir:2016zee} have discussed
whether the BH's spin constrains the binary evolution path
in the case of Pop I and Pop II binaries.
If we detect a lot of Pop III BBH mergers, we might be able to constrain
the Pop III binary evolution paths not only by the mass distribution but also by the spin distribution, too.

One of the interesting output from the QNM GWs
is whether we can confirm the ergoregion of the Kerr BH. From Table~\ref{tab:detection_C},
the event rate for the confirmation of $>50\%$ of the ergoregion
is $0.040-3.1~{\rm events~yr^{-1}}~({\rm SFR_p}/ (10^{-2.5}~M_\odot~{\rm yr^{-1}~Mpc^{-3}})) 
\cdot ({\rm [f_b/(1+f_b)]/0.33})$ with SNR$>35$.

When we consider to extract the rotational energy of BHs
such as Penrose process~\cite{Penrose:1969pc}
and the Blanford-Znajek process~\cite{Blandford:1977ds},
we want to observe highly spinning remnant BHs.
For remnant BH's spins $q_f>0.95$,
the event rate with SNR$>35$ is 
$0.0027-0.037~{\rm events~yr^{-1}}~({\rm SFR_p}/ (10^{-2.5}~M_\odot~{\rm yr^{-1}~Mpc^{-3}})) 
\cdot ({\rm [f_b/(1+f_b)]/0.33})$ in the KAGRA detector from Table~\ref{tab:detection_all_q}.
A third-generation GW observatory,
the Einstein Telescope (ET)~\cite{Punturo:2010zz}
will have an improvement of the sensitivity by about a factor 10
from the second generations detectors.
This means that we have roughly 1000 times higher than
the current expected event rates,
and for example, the ringdown event rate with $q_f>0.95$
will become
\bea
3-30~{\rm events~yr^{-1}~\left(\frac{SFR_p} { 10^{-2.5}~M_\odot~yr^{-1}~Mpc^{-3}}\right) }
\cdot \left(\rm \frac{f_b/(1+f_b)}{0.33} \right) \left(\frac{\epsilon_r}{0.03}\right)^{1/2} \,.
\eea
Here, we have introduced $\epsilon_r$ as
the fraction of the BH mass radiated in the ringdown phase,
and assumed $\epsilon_r=3\%$ to calculate the SNR and the event rates in this paper.
If $\epsilon_r=0.3\%$, we will still have a possibility to detect the QNM GWs from
highly spinning remnant BHs.

Finally, Pop III BBH mergers is possible to be a target 
for space-based GW detectors
such as eLISA~\cite{Seoane:2013qna} and DECIGO~\cite{Seto:2001qf}.
The study in this direction is one of our future works.

%%%%%%%%%%%%%%%%%%%%%%%%%%%%%%%%%%%%%%%%
\section*{Acknowledgment}
%%%%%%%%%%%%%%%%%%%%%%%%%%%%%%%%%%%%%%%%

This work was supported by MEXT Grant-in-Aid for Scientific Research
on Innovative Areas,
``New Developments in Astrophysics Through Multi-Messenger Observations
of Gravitational Wave Sources'', Nos.~24103001 and 24103006 (HN, TN, TT),
JSPS Grant-in-Aid for Scientific Research (C), No.~16K05347 (HN),
and the Grant-in-Aid from the Ministry of Education, Culture, Sports,
Science and Technology (MEXT) of Japan, No.~15H02087 (TN, TT).

%%%%%%%%%%%%%%%%%%%%%%%%%%%%%%%%%%%%%%%%
%\appendix

%%%%%%%%%%%%%%%%%%%%%%%%%%%%%%%%%%%%%%%%

%%%%%%%%%%%%%%%%%%%%%%%%%%%%%%%%%%%%%%%%
\end{document}